\documentclass[%
 reprint,
 amsmath,amssymb,
 aps,
]{revtex4-2}
\usepackage{dcolumn}
\usepackage[utf8]{inputenc}
\usepackage[T1]{fontenc}
\usepackage{url}
\usepackage{hyperref}
\usepackage{physics}
\usepackage{graphicx,lipsum}
\usepackage{amsmath}
\usepackage{amssymb}
\usepackage{amsfonts}
\usepackage{bbm}
\usepackage{dsfont}
\usepackage{color,soul}
\usepackage{cleveref}

\begin{document}

\preprint{APS/123-QED}

\title{Optimal Discrimination of Mixed Symmetric Multi-mode Coherent States}

\author{Ioannis Petrongonas}
\email{ip2004@hw.ac.uk}
%\altaffiliation[Also at ]{Physics Department, XYZ University.}
\author{Erika Andersson}
 \email{e.andersson@hw.ac.uk}
\affiliation{%
 Institute of Photonics and Quantum Sciences, Schoolf of Engineering and Physical Sciances, Heriot-Watt University, SUPA, Edinburgh
 %\\
}

%\collaboration{MUSO Collaboration}

\date{\today}

\begin{abstract}
%The idea of this paper is to establish how a specific way of mixing pure symmetric photon (coherent) states, by phase-randomizing the states,
We find the optimal measurement for distinguishing between symmetric multi-mode phase-randomized coherent states. A motivation for this is that phase-randomized coherent states can be used for quantum communication, including quantum cryptography. %, when these mixed states are used to encode classical information.
 %The main purpose of this specific work is to introduce a mathematical recipe of how can we optimally distinguish between those states, both for cheating and honest parties, which in general does not coincide with the square root measurement. The square root measurement, as we know, is the minimum error measurement for pure states and not for mixed states, as a whole. 
The so-called square-root measurement is optimal for pure symmetric states, but is not always optimal for mixed symmetric states. When phase-randomizing a multi-mode coherent state, the state becomes a mixture of pure multi-mode states with different total photon numbers.  %When distinguishing between phase-randomized coherent states, the 
We find that the optimal measurement for distinguishing between any set of phase-randomised coherent states can be realised by first counting the total number of photons, and then distinguishing between the resulting pure states in the corresponding photon-number subspace. If the multi-mode coherent states we started from are symmetric, then the optimal measurement in each subspace is a square-root measurement.
The overall optimal measurement in the cases we consider is also a square-root measurement. In some cases, we are able to present a simple linear optical circuit that realizes the overall optimal measurement.

%separates in the total photon number $N$ subspaces, which points out that in each one of these totally independent subspaces, we are dealing with pure states. So, as in the pure case scenario, we have to apply the square root measurement, for each one of the photon number subspaces, in order to derive the optimal measurement result that we desire.%

\end{abstract}

\maketitle

%\begin{center}
%\textbf{Introduction}
%\end{center}
\section{Introduction}
%The subject of 
Quantum state discrimination has attracted
a lot of interest in the past 50 years. %It has been shown that 
Orthogonal quantum states, both mixed and pure, can be perfectly distinguished from each other, but for non-orthogonal states this is impossible even in principle. ``Imperfect" state discrimination is still possible, and can be optimised in many different ways~\cite{Barnett2008,Bergou2004}. %There are many ways to define what an ``optimal" measurement is.
In this paper, we will be concerned with finding minimum-error measurements for phase-randomized multi-mode coherent states. There is a definite phase relationship between the relative phases of the coherent states in the different modes, but the overall phase is random. This is a situation that arises for example when a sender and receiver do not share a common phase reference, including when noise effectively randomises the overall phase.
In a minimum-error measurement, one minimises the probability that the result is wrong. A second type of optimal measurements which will be relevant is unambiguous or ``error-free" measurements. This type of measurement is possible for some sets of states; the results are then guaranteed to be correct, at the expense of including an inconclusive ``failure" outcome. 
%\Erika{Other types of optimal quantum measurements include maximum mutual information measurements and maximum confidence measurements {\bf include citations; it sounded almost as if minimum error and unambiguous are the only options. Alternatively, remove the remark about mutual info and max conf but state more clearly that min err and unambiguous are just two options among others}.} \Ioannis{We can write some text that minimum-error and unambiguous measurements are two types from a big variety of options without specifying to the mutual info and max confidence ones.}

Section \ref{sec:background} is a brief review of minimum-error quantum measurements, including the so-called square-root measurement, and unambiguous quantum measurements.
In Section \ref{sec:optdisc}, we find the optimal measurement for distinguishing between equiprobable symmetric phase-randomized multi-mode coherent states,
and look at some examples of this general result for two, three and four modes. In Section \ref{sec:OT}, we apply the results to a quantum protocol for 1-out-of-2 oblivious transfer, where one needs the probability for distinguishing between four mixed symmetric states, and to quantum retrieval games. Finally, we discuss how our results contribute to the field of mixed-state discrimination, and for understanding practical ways for distinguishing between quantum states in communication protocols.
%{\bf Revise the preceding sentence when the Discussion is finalised.}

\section{Minimum-error and unambiguous quantum measurements}
\label{sec:background}

In a minimum-error measurement, the probability that the obtained result is correct is as high as possible.
Holevo and Helstrom~\cite{Holevo1973, Helstrom1976} obtained the minimum-error measurement for distinguishing between any two quantum states, pure or mixed. For more than two states, it is generally difficult to find optimal measurements. %and it depends on the type of states we deal with. 
One exception is for symmetric pure states, where the minimum-error measurement is known to be the so-called square-root or ``pretty good" measurement~\cite{Barnett2008,Barnett2001,ChouHsu2003,EldarMegreVerg2004}.

Unambiguous state discrimination means that  an inconclusive outcome is permitted, but the rest of the results are guaranteed to be correct. %, has also been investigated for a number of cases. 
The goal is to minimise the probability of the inconclusive result. %, which is the same as maximizing the probability for a conclusive result. 
%On the other hand, numerous cases have been investigated when the possibility of an inconclusive outcome to occur is permitted, meaning the unambiguous state discrimination. 
For two equiprobable pure states, unambiguous discrimination was analysed by Ivanovic~\cite{Ivanovic1987}, Dieks~\cite{Dieks1988}, and Peres~\cite{Peres1988}.  %Ivanovic, Dieks and Peres analysed the case where the a priori probabilities of the two pure states are equal, while 
Jaeger and Shimony~\cite{Jaeger&Shimony1995} solved the case for unequal a priori probabilities. For more than two pure states, it was shown by Chefles~\cite{Chefles1998} that unambiguous discrimination with a nonzero success probability (for every possible state) is only possible if the states are linearly independent. 
%\Erika{\bf I agree with the maths in this paper, but don't agree that this means that unambiguous discrimination is possible only for linearly independent states. I'll explain. We might rephrase the previous sentence to make it OK.} 
Only for some specific cases, such as symmetric pure states~\cite{CheflesSym1998}, has there been an analytical solution for the optimal unambiguous measurement for $N$ linearly independent pure states.
%\Erika{\bf Is this for an unambiguous measurement?} \Ioannis{Yes, Chefles derives the maximum attainable value of the unambiguous discrimination probability $P_D$ for symmetric states with equal a priori probabilities. We might need to rephrase the previous line.}  
For mixed states, the optimal unambiguous measurement is also only known in some special cases
%again some limited examples have been solved~\cite{Rudolph2003,SunBergouHillery2002,BergouHerzogHillery2003,HerzogBergou2005,Raynal2007,HerzogUl2007,HerzogMaxCon2008}. 
%\Erika{\bf What types of measurements are the preceding ones? From the text it sounds as if they are all for unambiguous measurements. If not, we need to rephrase.} \Ioannis{All of the references $17-23$ correspond to umambiguous state discrimination.} 
Lower and upper bounds for the probability of an inconclusive result have been derived  for $N$ pure or mixed states~\cite{Rudolph2003,FengDuanYing2004,RaynalLowBou2005,Fiur_ek_2003,Montanaro2008}.
Numerical optimisation, such as semidefinite programming, can also be used to find optimal measurements both for unambiguous~\cite{Sun2002, Eldar2003} and minimum-error measurements~\cite{EldarMegreVerg2003}. %\Ioannis{The first two citations are for unambiguous measurement but the third citation is for ambiguous state discrimination.}

\section{Optimal discrimination of phase-randomized coherent states}
\label{sec:optdisc}

Phase-encoded coherent states are frequently used to encode information in quantum communication~\cite{Sych_2010,Cao_2015,Lo2005PhaseRI,van_Enk_2002,Grosshans_2002}. We will consider how to distinguish between phase-randomized multi-mode coherent states. This corresponds to a situation where the receiver (who could be a legitimate receiver, or perhaps an adversary in a quantum cryptographic protocol) and sender share no strong phase reference. Instead, information is encoded in relative phase differences between different modes.

In order to construct the states we will distinguish between, we start from different sets of symmetric pure multi-mode coherent states. A set of $L$ equiprobable states $\{\ket{\psi_K}\}$, with $K=0,1,\ldots,L-1$, is said to be (singly) symmetric if there is a unitary operation $U$ for which it holds that $U^L=\mathds{1}$, and $\ket{\psi_K}=U^K\ket{\psi_0}$.
When phase-randomizing pure symmetric states they remain symmetric, but become mixed. %, as is also explicitly shown in Appendix \ref{app:general}. 

For distinguishing between two or more states $\rho_i$, each occurring with probability $p_i$, the so-called {\em square-root measurement} or {\em pretty good measurement} is optimal in some  cases, in particular for equiprobable symmetric pure states~\cite{Barnett2008, Barnett2001, ChouHsu2003, EldarMegreVerg2004}. The measurement operators for this measurement are $\pi_i=\rho^{-\frac{1}{2}}p_i\rho_i\rho^{-\frac{1}{2}}$, where $\rho=\sum_ip_i\rho_i$. The corresponding success probability is
\begin{equation} \label{eq:squarerootmeasurement}
P_\text{s}=\sum_ip_i\tr[\rho_i\big(\rho^{-\frac{1}{2}}p_i\rho_i\rho^{-\frac{1}{2}}\big)].
\end{equation}
The minimum-error measurement for pure symmetric states is the square-root measurement, but for mixed symmetric states, the square-root measurement is generally no longer optimal~\cite{ChouHsu2003,EldarMegreVerg2004,Wallden_2014}. It is therefore not immediately evident what the minimum-error measurement is for multi-mode phase-randomized coherent states. 

In Appendix \ref{app:general}, we prove that %if we start from a general set of pure symmetric states, phase-randomization leads to mixed symmetric states. Then, we show that 
each of the resulting phase-randomized mixed states is a mixture of pure states with different total photon number, with one pure state for each total photon number.  This holds whether or not the pure states we start from are symmetric or not. It follows that the minimum-error measurement for any set of multi-mode phase-randomized coherent states can be realised by first counting the total number of photons, followed by the minimum-error measurement that distinguishes between the resulting pure states in the relevant subspace. Moreover, if the pure multi-mode states we start from are symmetric, and the unitary symmetry operator $U$ does not change the total number of photons, then in each subspace with some total number of photons, the different possible pure states, each corresponding to a different $\rho_i$, are symmetric. The overall minimum-error measurement is then a combination of
the different square-root measurements for each photon-number subspace.
In Appendix \ref{app:general}, we explicitly show that this total measurement satisfies the conditions for the measurement to be optimal. 
Below, we give examples of optimal measurements for phase-randomized coherent states, obtained using the method outlined above.

\subsection{Two-mode coherent states} \label{num1}

We will first consider encoding one bit of classical information using phase-randomized versions of the two-mode coherent states
\begin{equation} \label{eq:twopure}
\ket{\psi_0}=\ket{\alpha,\alpha},\quad \ket{\psi_1}=\ket{\alpha,-\alpha}.
\end{equation}
The first mode can be thought of as a ``weak" phase reference of the same amplitude as the second mode; both modes will be phase-randomized, retaining their relative phase difference. A classical bit could be encoded in the relative phase difference of the first and second pulses. Before phase-randomization, the two states in eq. \eqref{eq:twopure} have the same pairwise overlap as $\ket{\alpha}$ and $\ket{-\alpha}$, and are just as distinguishable as these states. If we phase-randomize the states $\ket{\psi_0}$ and $\ket{\psi_1}$, %this corresponds to a situation where the communicating parties do not share any other phase reference, and 
the resulting states become less distinguishable.

Using the definition of a coherent state with amplitude $\alpha=\abs{\alpha}e^{i\theta}\in\mathbb{C}$ in the Fock or number state basis,
\begin{equation} \label{eq:coherent}
\ket{\alpha}=e^{-\frac{\abs{\alpha}^2}{2}}\sum_{j=0}^\infty\frac{\alpha^j}{\sqrt{j!}}\ket{j},
\end{equation}
we express the two-mode coherent states in eq. \eqref{eq:twopure} as
%\begin{equation} \label{eq3}
%\ket{\psi_0}=e^{-\abs{\alpha}^2}\sum_{j,k=0}^\infty\frac{\alpha^{j+k}}{\sqrt{j!k!}}\ket{j,k},\quad \ket{\psi_1}=e^{-\abs{\alpha}^2}\sum_{j,k=0}^\infty\frac{(-1)^k\alpha^{j+k}}{\sqrt{j!k!}}\ket{j,k}.
%\end{equation}
%We can also write the two states above as
\begin{equation} \label{eq:twopurecoherent}
\ket{\psi_b}=e^{-\abs{\alpha}^2}\sum_{j,k=0}^\infty\frac{(-1)^{bk}\alpha^{j+k}}{\sqrt{j!k!}}\ket{j,k},
\end{equation}
with $b=0,1$. By phase-randomizing the pure states $\ket{\psi_b}$, we obtain the mixed states
\begin{align} \label{eq:twomixed}
%\begin{split}
\rho_b &= \frac{1}{2\pi}\int_0^{2\pi}d\theta\ket{\psi_b}\bra{\psi_b} \nonumber\\
%&=&e^{-2\abs{\alpha}^2}\sum_{j,k,p,q=0}^\infty{\frac{(-1)^{b(k+q)}\abs{\alpha}^{j+k+p+q}}{\sqrt{j!k!p!q!}}\frac{1}{2\pi}\int_0^{2\pi}e^{i(j+k-p-q)\theta}d\theta
%\ket{j,k}\bra{p,q}}\nonumber \\
&=e^{-2\abs{\alpha}^2}\sum_{\substack{j,k,p,q=0\\ j+k=p+q}}^\infty{\frac{(-1)^{b(k+q)}\abs{\alpha}^{j+k+p+q}}{\sqrt{j!k!p!q!}}\ket{j,k}\bra{p,q}},
%\end{split}
\end{align}
where we used the integral representation of the Kronecker delta,
\begin{equation} \label{eq:kronecker}
\frac{1}{2\pi}\int_0^{2\pi}e^{i(j+k-p-q)\theta}d\theta=\delta_{j+k,p+q}.
\end{equation}
%We derive two types of measurement for ambiguous quantum state discrimination. We first describe the \textit{Helstrom measurement}.
In this first example, we are distinguishing between only two states.
The minimum-error measurement that optimally distinguishes between two quantum states, mixed or pure, is known to be the so-called {\em Helstrom measurement}~\cite{Barnett2008,Holevo1973,Helstrom1976}, obtained as follows. Given two states $\rho_0$ and $\rho_1$ with respective probabilities $p_0$ and $p_1$, we look for an optimal measurement with two measurement operators $\Pi_0$ and $\Pi_1$, corresponding to the two outcomes. From $\sum_i\Pi_i=\mathds{1}$, which ensures that the sum of the probabilities for all results is equal to 1, we have $\Pi_1=\mathds{1}-\Pi_0$. The probability that the obtained result is correct is given by
%for distinguishing between the two states is
\begin{equation} \label{eq:helstrom}
\begin{split}
P_\text{corr}&=p_0\tr(\rho_0\Pi_0)+p_1\tr(\rho_1\Pi_1) \\% & =p_0\tr(\rho_0\Pi_0)+p_1\tr[\rho_1(\mathds{1}-\Pi_0)] \\
& =p_1+\tr[(p_0\rho_0-p_1\rho_1)\Pi_0].
\end{split}
\end{equation}
To maximize the probability of obtaining a correct result $P_\text{corr}$, the trace of $(p_0\rho_0-p_1\rho_1)\Pi_0$ should be maximized. This is accomplished when $\Pi_0$ is a projector on the positive eigenspace of $p_0\rho_0-p_1\rho_1$; $\Pi_1$ is then a projector on the negative eigenspace of this same operator. If there are eigenvalues equal to zero, the corresponding eigenvectors can be included either in $\Pi_0$ or in $\Pi_1$, or a random guess can be made in this case, without affecting $P_\text{corr}$. 

The Helstrom measurement operators for the two-mode phase-randomized states in \eqref{eq:twomixed} 
%$\Pi_i$ and the states $\rho_i'$ for $i=0,1$ 
are derived in Appendix \ref{app:general} (see \cref{eq:optimalmeasurementphaserandomizedtwomodecase1,eq:optimalmeasurementphaserandomizedtwomodecase2}). In this particular case, however, the optimal measurement can alternatively be both derived and experimentally realised by considering that one can interfere the states on a balanced beam splitter, acting  as $U_{\rm BS}|\alpha, \beta\rangle = |(\alpha+\beta)/\sqrt 2, (\alpha-\beta)/\sqrt 2\rangle$, which transforms the incoming coherent states $\ket{\alpha,\alpha}$ and $\ket{\alpha,-\alpha}$ according to
\begin{equation} \label{eq:beamsplitter}
\begin{split}
U_\text{BS}\ket{\psi_0}& =U_\text{BS}\ket{\alpha,\alpha}=|\sqrt{2}\alpha,0\rangle,\\ 
U_\text{BS}\ket{\psi_1}& =U_\text{BS}\ket{\alpha,-\alpha}=|0,\sqrt{2}\alpha\rangle .
\end{split}
\end{equation}
When phase-randomizing the states in \cref{eq:beamsplitter}, they become mixtures of states with different photon numbers in the two output modes. If the input state was $\ket{\alpha,\alpha}$, photons exit only in the first mode, and if it was $\ket{\alpha,-\alpha}$, in the second output mode. Phase randomization does not affect how many photons exit the beam splitter in each mode, and it also does not matter whether phase-randomization takes place before or after the beam splitter. This means that after applying $U_{\rm BS}$ and phase randomization, the optimal measurement is to detect which output contains photons. If no photons are detected, one guesses the state which was most likely. This measurement is closely related to the optimal unambiguous measurement, since detecting photons in one of the outputs unambiguously identifies the state. The difference is that for the optimal unambiguous measurement, if no photons are detected, one records a ``failure" outcome. 

\begin{figure*}
\centering
\includegraphics[scale=0.7]{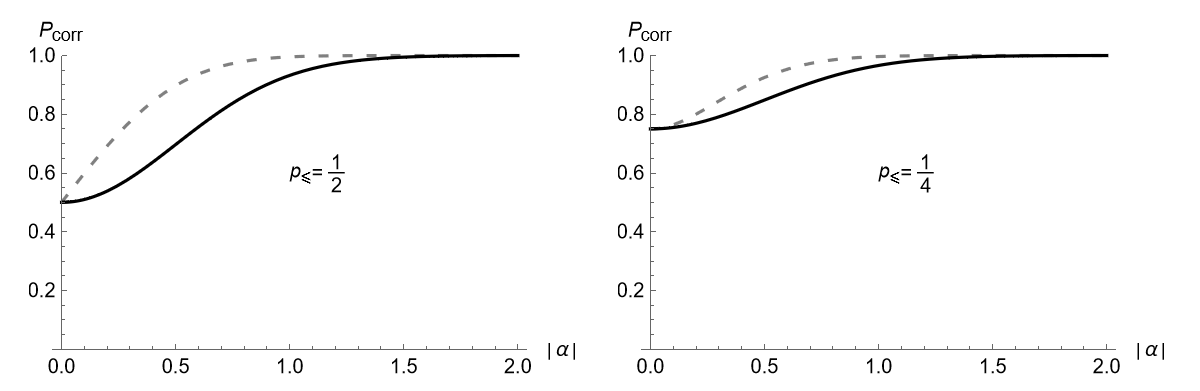}
\caption{The probability $P_{\text{corr}}$ for the minimum-error measurement result to be correct, as a function of $\abs{\alpha}$, for the pure two-mode states in \cref{eq:twopure} (dashed grey lines) and the mixed two-mode states states in \eqref{eq:twomixed} (solid black lines), for equiprobable states, with $p_\leqslant=\frac{1}{2}$ (left) and for $p_\leqslant=\frac{1}{4}$ (right).} 
\label{fig1}
\end{figure*}

The resulting Helstrom probability for the minimum-error measurement result to be correct %for the $2$-mode phase-randomized states in \eqref{eq:twomixed} 
is
\begin{equation}
\label{eq:twomixedhelstrom}
P_{\text{corr}}=%p_0\Tr(\Pi_0\rho_0')+p_1\Tr(\Pi_1\rho_1')=
1-p_\leqslant e^{-2\abs{\alpha}^2},
\end{equation}
where $p_\leqslant=\min(p_0,p_1)$. % for $p_0\neq p_1$.  and $p_\leqslant=1/2$ for $p_0=p_1$. 
For the corresponding pure states in \cref{eq:twopure}, the Helstrom measurement is correct with probability
\begin{equation}
\label{eq:twopurehelstrom}
P_{\text{corr}}^{\text{pure}}=\frac{1}{2}+\frac{1}{2}\sqrt{1-4p_\leqslant(1-p_\leqslant)e^{-4\abs{\alpha}^2}}.
\end{equation}
We plot the Helstrom probabilities both for pure and phase-randomized states in \cref{fig1}. It can be seen that as expected, the probability to correctly identify the state is higher for pure states than for the corresponding mixed states.
%\Erika{\bf It would be good if this plot can also show the Helstrom success probability for the states before phase randomization (for pure states). This way one sees how much lower the success probability becomes due to phase randomization.}

Interestingly, the Helstrom measurement (the optimal minimum-error measurement for two states) can for the phase-randomized two-mode states be experimentally realised relatively easily. This is in contrast to the minimum-error measurement for distinguishing between the pure states $\ket{\alpha}$ and $\ket{-\alpha}$, or equivalently, between the pure states $\ket{\alpha, \alpha}$ and $\ket{\alpha,-\alpha}$, where experimental realisation is much less straightforward~\cite{Dolinar,Kennedy,Takeoka_2008}. The optimal unambiguous measurement for distinguishing between the pure states $\ket{\alpha}$ and $\ket{-\alpha}$ is also realized by interfering the state on a balanced beam splitter with a state $\ket{\alpha}$, and detecting photons at the outputs; no photons corresponds to the ``failure" outcome. The optimal unambiguous measurements for the states $\ket{\alpha}$ and $\ket{-\alpha}$, for the states $\ket{\alpha,\alpha}$ and $\ket{\alpha,-\alpha}$, and for the phase-randomized two-mode states in equation \eqref{eq:twomixed}, all have the same success and failure probabilities. That is, phase randomization does in this case not increase the failure probability for an unambiguous measurement, while decreasing the success probability for a minimum-error measurement. This could be useful for protocols in quantum cryptography, if an honest party should perform an unambiguous measurement, while a dishonest party might cheat using a minimum-error measurement. Phase randomization would then not increase the failure probability for an honest party, while decreasing the cheating probability for a dishonest party. 
%\Erika{\bf Can move that discussion to the ``Discussion" at the end if desired. Note: phase randomization of the coherent states $\ket{\alpha}$ and $\ket{-\alpha}$ does not appear useful for Rabin oblivious transfer, where Alice sends Bob one of two non-orthogonal states; an honest Bob makes an unambiguous measurement that succeeds with some probability. A cheating Bob should make a minimum-error measurement, and phase randomization lowers his cheating probability. However, since Alice has complete control over the whole state that goes to Bob, she can perfectly steer his outcome to her advantage and cheat with probability 1. If we can't find an example where phase randomization of two-mode states is useful for quantum crypto, do we not mention it? Somebody else might still find an example.}

For symmetric mixed states, the square-root measurement is generally not optimal, but in Appendix \ref{app:twomode} , we also show that if the two-mode phase-randomized states in \eqref{eq:twomixed} are equiprobable, then the square-root measurement is the same as the minimum-error measurement. 
%\Erika{\bf We might also consider a section discussing when the square-root measurement is and isn't optimal for symmetric mixed states, and include that discussion there. This discussion quite likely would not appear exactly here, but we would refer to it here.}%

\subsection{Three-mode coherent states} \label{num2}

Next, we consider the four pure three-mode coherent states %, which encode two classical bits of information
\begin{align} \label{eq:purethreemode}
&\ket{\psi_{00}}=\ket{\alpha,\alpha,\alpha},
\quad\quad~
\ket{\psi_{01}}=\ket{\alpha,\alpha,-\alpha},\nonumber\\
&\ket{\psi_{11}}=\ket{\alpha,-\alpha,-\alpha},\quad \ket{\psi_{10}}=\ket{\alpha,-\alpha,\alpha},
\end{align}
which can encode two bits of classical information. These states are coherent-state versions of the qutrit states used in a protocol for XOR quantum oblivious transfer~\cite{Stroh_2023}, 
%\begin{equation}
%\label{eq:qutritstates}
%\begin{split}
%\ket{\psi_{00}^\rm trit}&=\frac{1}{\sqrt 3}(\ket 0+ \ket 1 +\ket 2), \\
%\ket{\psi_{01}^\rm trit}&=\frac{1}{\sqrt 3}(\ket 0+ \ket 1 -\ket 2), \\
%\ket{\psi_{11}^\rm trit}&=\frac{1}{\sqrt 3}(\ket 0- \ket 1 -\ket 2), \\
%\ket{\psi_{10}^\rm trit}&=\frac{1}{\sqrt 3}(\ket 0- \ket 1 +\ket 2),
%\end{split}
%\end{equation}
\begin{equation}
\label{eq:qutritstates}
\begin{split}
\ket{\psi_{00}^{\rm trit}} &= \frac{1}{\sqrt{3}}(\ket{0} + \ket{1} + \ket{2}), \\
\ket{\psi_{01}^{\rm trit}} &= \frac{1}{\sqrt{3}}(\ket{0} + \ket{1} - \ket{2}), \\
\ket{\psi_{11}^{\rm trit}} &= \frac{1}{\sqrt{3}}(\ket{0} - \ket{1} - \ket{2}), \\
\ket{\psi_{10}^{\rm trit}} &= \frac{1}{\sqrt{3}}(\ket{0} - \ket{1} + \ket{2}).
\end{split}
\end{equation}
and would be natural to consider if one wants to construct a corresponding protocol using coherent states instead of single qutrits.
The states in \cref{eq:purethreemode} are both singly and doubly symmetric. They are singly symmetric since it holds that
\begin{equation} \label{eq:singlysymmetric}
\begin{split}
\ket{\psi_{00}}&,\quad \ket{\psi_{01}}=U\ket{\psi_{00}}, \\ \ket{\psi_{11}}&=U^2\ket{\psi_{00}},\quad \ket{\psi_{10}}=U^3\ket{\psi_{00}}
\end{split}
\end{equation}
with the unitary transformation 
\begin{align} \label{eq:unitarythreemode}
U&=\sum_{j,k,l=0}^\infty(-1)^l\ket{j,k,l}\bra{j,l,k} \notag \\
&=\bigoplus_{N=0}^\infty\sum_{\substack{j,k,l=0\\ j+k+l=N}}^N(-1)^l\ket{j,k,l}\bra{j,l,k}=\bigoplus_{N=0}^\infty U_N
\end{align}
obeying $U^4=\mathds{1}$. In the expression above, we also decomposed $U$ in terms of unitary transforms $U_N$, acting in subspaces with different total photon number $N$, each one also obeying $U^4_N=\mathds{1}$ in its respective subspace. The states are also doubly symmetric, since they can be expressed as
%\begin{equation} \label{eq:doublysymmetric}
%\ket{\psi_{00}},\ \ \ket{\psi_{01}}=V\ket{\psi_{00}},\ \ \ket{\psi_{11}}=WV\ket{\psi_{00}}=VW\ket{\psi_{00}},\ \ \ket{\psi_{10}}=W\ket{\psi_{00}}
%\end{equation}
\begin{equation} \label{eq:doublysymmetric}
\ket{\psi_{ij}}=W^iV^j\ket{\psi_{00}},\quad i,j=0,1
\end{equation}
where the unitaries 
\begin{equation}
\label{eq:doublysymmetricunitaries}
\begin{split}
V&=\sum_{j,k,l=0}^\infty(-1)^l\ket{j,k,l}\bra{j,k,l}, \\ W&=\sum_{j,k,l=0}^\infty(-1)^k\ket{j,k,l}\bra{j,k,l}
\end{split}
\end{equation}
satisfy $V^2=W^2=\mathds{1}$ and $[V,W]=0$. Evidently, the above four pure three-mode coherent states have the same pairwise overlaps as the four two-mode coherent states 
%\begin{equation} \label{eq:twomodesameoverlap}
%\ket{\psi_{00}'}=\ket{\alpha,\alpha},\ \ \ket{\psi_{01}'}=\ket{\alpha,-\alpha},\ \ \ket{\psi_{11}'}=\ket{-\alpha,-\alpha},\ \ \ket{\psi_{10}'}=\ket{-\alpha,\alpha}.
%\end{equation}
\begin{equation}
\label{eq:twomodesameoverlap}
\ket{\psi_{ij}'}=\ket{(-1)^i\alpha,(-1)^j\alpha},\quad i,j=0,1
\end{equation}
The first mode in the three-mode states %which is not taking part in the shifting and phase factor operations, 
can again be thought of as a phase reference. 
Phase randomization of the three-mode states gives
%means that there exists one global unitary transformation $W$ that connects these four states (permute modes 2 and 3 and add a phase shift either to the new mode 2 or the new mode 3)%
%that this set of states can be obtained by repeated application of two commuting unitary transforms $U$ and $V$ (add a phase shift -1 either to the second mode or to the third mode).%
%if we neglect the first of the three modes, which is common for all of them.% 
\begin{equation} \label{eq:threemodemixed}
\begin{split}
\rho_{bc} & =\frac{1}{2\pi}\int_0^{2\pi}d\theta\ket{\psi_{bc}}\bra{\psi_{bc}}\\
&=e^{-3\abs{\alpha}^2}\sum_{\substack{j,k,l,\\ p,q,r=0\\ j+k+l=\\p+q+r}}^\infty{\frac{(-1)^{\mathcal{J}_{bc}}\abs{\alpha}^{2(j+k+l)}}{\sqrt{j!k!l!p!q!r!}}\ket{j,k,l}\bra{p,q,r}},
\end{split}
\end{equation}
where 
\begin{equation} \label{eq:threemodeindex}
\mathcal{J}_{bc}=b(k+q)+c(l+r),
\end{equation}
with $b,c=0,1$. The states $\rho_{bc}$ are expressed in terms of states with different total photon numbers $N=j+k+l=p+q+r$, and the $\rho_{bc}$ are evidently ``block diagonal", with the blocks corresponding to subspaces with different total photon number. The states $\rho_{bc}$ can in fact be written as mixtures of pure states, each one with different total photon number (see Appendix \ref{app:general}),
\begin{align} \label{eq:mixturepuresymmetricstates}
\rho_{bc}&=\bigoplus_{N=0}^\infty{p_N \ket{\psi_{bc,N}}\bra{\psi_{bc,N}}
%\rho_{bc,N}
} \notag \\
&=\bigoplus_{N=0}^\infty{e^{-3\abs{\alpha}^2}M_N\abs{\alpha}^{2N}\ket{\psi_{bc,N}}\bra{\psi_{bc,N}}},
\end{align}
where $b,c=0,1$, $p_N$ is the probability for the state to contain $N$ photons, and the pure states $\ket{\psi_{bc,N}}$ are given by
\begin{equation} \label{eq:threemodeNpure}
\ket{\psi_{bc,N}}=\frac{1}{\sqrt{M_N}}\sum_{\substack{j,k,l=0\\ j+k+l=N}}^N{\frac{(-1)^{bk+cl}}{\sqrt{j!k!l!}}\ket{j,k,l}},
\end{equation}
where $M_N={3^N}/{N!}$ is a normalization factor. Pure states with different photon numbers $N$ and $N'$ are of course orthogonal,
%\begin{equation} \label{eq:orthogonalitysubspaces}
$\bra{\psi_{bc,N}}\ket{\psi_{de,N'}}=0.$
%\\ \text{or}\ \ \rho_{ab,N}\rho_{cd,N'}=0.
%\end{equation}
Each total density matrix $\rho_{bc}$ in \cref{eq:threemodemixed} is therefore a mixture of perfectly distinguishable pure states with different photon numbers.

%After phase randomization, we have four pure symmetric states in each total photon number subspace. It holds that%
Because both $\rho_{bc}$ and $U$ can be decomposed in terms of photon-number subspaces, and we have $U_N^4=\mathds{1}$ in each subspace,
%it holds that
%\begin{equation} \label{eq20}
%U\rho_{bc}U^\dag=\sum_{N=0}^\infty p_NU_N\ket{\psi_{bc,N}}\bra{\psi_{bc,N}}U_N^\dag,
%\end{equation}
%with $(U_N)^4=\mathds{1}$. It is clear from \eqref{eq20} that for the four symmetric total mixed states $\rho_{bc}$, 
the four resulting pure states $\ket{\psi_{bc,N}}$ in each total photon-number subspace are also symmetric. It follows that the optimal measurement that distinguishes between the three-mode phase-randomized multi-mode coherent states $\rho_{bc}$ is to make a projection onto the different subspaces with different total photon numbers, followed by a square-root measurement for distinguishing between the pure symmetric states in the resulting subspace. The square-root measurements are different in each subspace; however, collectively the overall measurement is the square-root measurement for the mixed states $\rho_{bc}$.

\begin{figure}[h]
\centering
\includegraphics[scale=0.7]{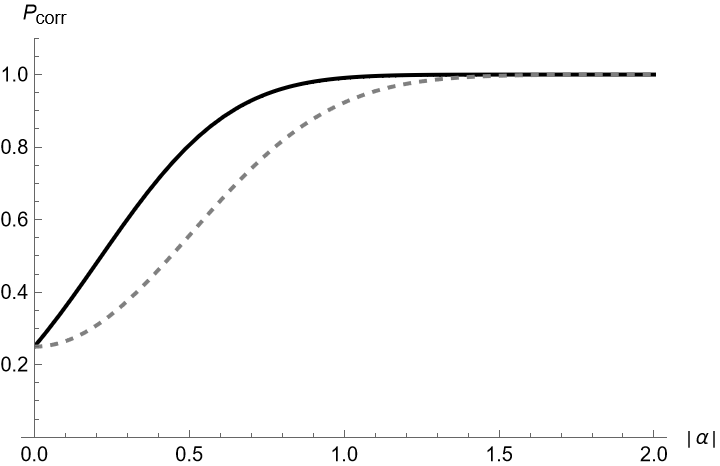}
\caption{{
Minimum-error} probability $P_{\text{corr}}$ as a function of $\abs{\alpha}$ for the equiprobable pure three-mode coherent states $\ket{\alpha, \alpha, \alpha}, \ket{\alpha, \alpha, -\alpha}, \ket{\alpha, -\alpha, -\alpha}, \ket{\alpha, -\alpha, \alpha}$ (solid black line) and their phase-randomized counterparts (dashed grey line). As expected, phase-randomization lowers the probability that the measurement result is correct. 
%\Ioannis{For small $\abs{\alpha}$, the coherent states with or without phase randomization can achieve less $B_{\text{OT}}$. For large $\abs{\alpha}$, the four symmetric coherent states become perfectly distinguishable and the qutrit states perform better in terms of the cheating probability. What type of comparison do you mean Erika ?}
}
\label{fig2}
\end{figure}

We will now derive the success probabilities for the optimal measurements in each subspace, and the overall success probability.
%For a set of equiprobable pure symmetric states, the optimal minimum-error measurement is the so-called square-root measurement~\cite{Barnett2008}. 
The success probability for optimally distinguishing between pure symmetric states using a square-root measurement can be obtained in terms of the
sum of the square roots of the eigenvalues of the Gram matrix~\cite{Wallden_2014}. The elements of the Gram matrix for a set of states
$\{\ket{\psi_i}\}$ are given by $\mathcal{G}_{ij}=\bra{\psi_i}\ket{\psi_j}$. We denote the pairwise overlaps between the four symmetric pure states in the subspace with $N$ photons by $F_N=\bra{\psi_{00,N}}\ket{\psi_{01,N}}=\bra{\psi_{01,N}}\ket{\psi_{11,N}}=\bra{\psi_{11,N}}\ket{\psi_{10,N}}=\bra{\psi_{10,N}}\ket{\psi_{00,N}}$ and $G_N=\bra{\psi_{00,N}}\ket{\psi_{11,N}}=\bra{\psi_{01,N}}\ket{\psi_{10,N}}$.
The Gram matrix for the four pure symmetric states in \cref{eq:threemodeNpure} is then (see Appendix \ref{app:threemode})
\begin{equation} \label{eq:grammatrixNthreemode}
\mathcal{G}_N=%\begin{pmatrix}
%1 & F_N & G_N & F_N^* \\
%F_N^* & 1 & F_N & G_N \\
%G_N & F_N^* & 1 & F_N \\
%F_N & G_N & F_N^* & 1
%\end{pmatrix}=
\begin{pmatrix}
1 & (\frac{1}{3})^N & (-\frac{1}{3})^N & (\frac{1}{3})^N \\
(\frac{1}{3})^N & 1 & (\frac{1}{3})^N & (-\frac{1}{3})^N \\
(-\frac{1}{3})^N & (\frac{1}{3})^N & 1 & (\frac{1}{3})^N \\
(\frac{1}{3})^N & (-\frac{1}{3})^N & (\frac{1}{3})^N & 1
\end{pmatrix}.
%=\frac{1}{3^N}\begin{pmatrix}
%3^N & 1 & (-1)^N & 1 \\
%1 & 3^N & 1 & (-1)^N \\
%(-1)^N & 1& 3^N & 1\\
%1 & (-1)^N & 1 & 3^N
%\end{pmatrix}
\end{equation}
The probability of correctly identifying one of the four equiprobable pure symmetric states with $N$ photons, in terms of the eigenvalues of the above Gram matrix~\cite{Wallden_2014,OT2021}, becomes
\begin{align} \label{eq:probcorrNthreemode}
P_{\text{corr},N}&=\frac{1}{4^2}\abs{\sum_{i=0}^3\sqrt{\lambda_i^{(N)}}}^2  \\
&=\frac{1}{16}\Big[3\sqrt{1-(-3)^{-N}}+\sqrt{1-(-3)^{-N+1}}\Big]^2. \notag
\end{align}
The total probability of correctly identifying one of the four phase-randomized three-mode coherent states $\rho_{bc}$ in \cref{eq:threemodemixed} then equals
\begin{alignat}{3}
\label{eq:probcorrthreemode}
P_{\text{corr}}&=\sum_{N=0}^\infty p_NP_{\text{corr},N} \notag \\
&=\frac{e^{-3\abs{\alpha}^2}}{16}\sum_{N=0}^\infty\frac{(3\abs{\alpha}^2)^N}{N!}&&\Big[3\sqrt{1-(-3)^{-N}} %\notag 
\\ & &&+\sqrt{1-(-3)^{-N+1}}\Big]^2.&&\notag
\end{alignat}
The probability to correctly distinguish between the respective pure three-mode coherent states in \cref{eq:purethreemode} is given by (see Appendix \ref{app:protsuccheatprob})
\begin{equation}
\label{eq:probcorrpurethreemode}
P_{\text{corr}}^{\text{pure}}=\frac{1}{4}\big(1+\sqrt{1-e^{-4\abs{\alpha}^2}}\big)^2.
\end{equation}
In \cref{fig2}, we plot the probability that the minimum-error measurement is correct for the phase-randomized three-mode states, $P_{\rm corr}$ in \cref{eq:probcorrthreemode}, and for the corresponding pure states before phase randomization, $P_{\rm corr}^{\rm pure}$ in \cref{eq:probcorrpurethreemode}, as a function of $|\alpha|$. As expected, phase randomization decreases the probability that the measurement result is correct. The minimum-error probability is equal to $1/4$ when $|\alpha|=0$, corresponding to a random guess when the states are indistinguishable, and increases asymptotically to 1 when $|\alpha| \rightarrow \infty$. For comparison, the minimum-error probability for the qutrit states in equation \eqref{eq:qutritstates} is equal to $\frac{3}{4}$. For coherent states, depending on $|\alpha|$, the minimum-error probability can be either smaller than or larger than this.

\subsection{Four-mode coherent states} \label{num3}

We then investigate the optimal measurement for phase-randomized versions of the four-mode coherent states $|\tilde{\psi}_{00}\rangle=\ket{\alpha,\alpha,\alpha,-\alpha}$, $|\tilde{\psi}_{01}\rangle=\ket{\alpha,\alpha,-\alpha,\alpha}$, $|\tilde{\psi}_{11}\rangle=\ket{\alpha,-\alpha,\alpha,\alpha}$ and $|\tilde{\psi}_{10}\rangle=\ket{-\alpha,\alpha,\alpha,\alpha}$. Again, these states are evidently symmetric, with a symmetry operation that permutes the modes so that $4\rightarrow 3, 3\rightarrow 2, 2\rightarrow 1$, and $1\rightarrow 4$. This unitary symmetry operator %for the four pure symmetric states $|\tilde{\psi}_{bc}\rangle$ is
can be written as
\begin{equation} \label{eq:fourmodeunitary}
\tilde{U}=\sum_{j,k,l,m=0}^\infty\ket{j,k,l,m}\bra{m,j,k,l}.
\end{equation}
The corresponding four phase-randomized states are
\begin{alignat}{3} \label{eq:fourmodemixed}
\tilde{\rho}_{bc} & =\frac{1}{2\pi} \int_0^{2\pi}d\theta|\tilde{\psi}_{bc}\rangle\langle\tilde{\psi}_{bc}| \notag \\
&=e^{-4\abs{\alpha}^2}\sum_{\substack{j,k,l,m,\\ p,q,r,s=0\\ j+k+l+m=\\p+q+r+s}}^\infty&&\frac{(-1)^{\tilde{\mathcal{J}}_{bc}}\abs{\alpha}^{2(j+k+l+m)}}{\sqrt{j!k!l!m!p!q!r!s!}}\times \notag \\
& &&\times
\ket{j,k,l,m}\bra{p,q,r,s},&&
\end{alignat}
where 
\begin{equation} \label{eq:fourmodeindex}
\tilde{\mathcal{J}}_{bc}=\bar{b}\bar{c}(m+s)+\bar{b}c(l+r)+bc(k+q)+b\bar{c}(j+p),
\end{equation}
with $\bar{b}=b+1$ (addition modulo $2$).
The phase-randomized states $\tilde\rho_{bc}$ can also be written as mixtures of pure states with different total photon number,
\begin{equation}
\tilde\rho_{bc} = \bigoplus_{N=0}^\infty \tilde p_N |\tilde\psi_{bc,N}\rangle\langle\tilde\psi_{bc,N}|,
\end{equation}
where
%we define %consider 
%the four pure symmetric states
\begin{equation} \label{eq:fourmodeNpure}
|\tilde{\psi}_{bc,N}\rangle=\frac{1}{\sqrt{\tilde{M}_N}}\sum\limits_{\substack{j,k,l,m=0\\ j+k+l+m=N}}^N{\frac{(-1)^{\mathcal{S}_{bc}}}{\sqrt{j!k!l!m!}}\ket{j,k,l,m}},
\end{equation}
$\tilde p_N = e^{-4\abs{\alpha}^2}\tilde{M}_N\abs{\alpha}^{2N}$, and
\begin{equation} \label{eq:fourmodeNpureindex}
\mathcal{S}_{bc}=\bar{b}\bar{c}m+\bar{b}cl+bck+b\bar{c}j,
\end{equation}
and $\tilde{M}_N={4^N}/{N!}$.

The unitary symmetry operation again does not alter the total photon number, so that it can be decomposed in different symmetric unitary operations, each acting in one total photon number subspace. Therefore, in each photon-number subspace, the states $|\tilde\psi_{bc,N}\rangle$ are symmetric, with $bc = 00, 01, 10, 11$. The optimal minimum-error measurement for distinguishing between the phase-randomized states $\tilde\rho_{bc}$ will again be to first project on the total photon number, followed by the square-root measurement for distinguishing between the states $|\tilde\psi_{bc,N}\rangle$ in the resulting photon number subspace. The states $|\tilde\psi_{bc,N}\rangle$ are in this case actually orthonormal for different choices of $b,c$, and therefore we can perfectly distinguish between the phase-randomized states $\tilde\rho_{bc}$ if we see at least one photon in total. If we see no photons, then we have to make a guess, which is correctly with probability $1/4$.
%We will again use the Gram matrices to work out the minimum-error probabilities for each total number of photons. For $N>0$, the Gram matrix for the states $|\widetilde{\psi}_{bc,N}\rangle$ is
%\begin{equation} %\label{eq:grammatrixfourmodeNpositive}
%\widetilde{\mathcal{G}}_N=\mathds{1}=\begin{pmatrix}
%1 & 0 & 0 & 0 \\
%0 & 1 & 0 & 0 \\
%0 & 0 & 1 & 0 \\
%0 & 0 & 0 & 1
%\end{pmatrix}.
%\end{equation}
%The Gram matrix for $N=0$ (zero total number of photons) is
%\begin{equation} %\label{eq:grammatrixfourmodeNzero}
%\widetilde{\mathcal{G}}_0=\begin{pmatrix}
%1 & 1 & 1 & 1 \\
%1 & 1 & 1 & 1 \\
%1 & 1 & 1 & 1 \\
%1 & 1 & 1 & 1
%\end{pmatrix},
%\end{equation}
%which has eigenvalues $\tilde{\lambda}_0^{(0)}=4$ and $\tilde{\lambda}_{1,2,3}^{(0)}=0$. 
%The probability to be correct for the subspace with $N=0$ is $\tilde{P}_{\text{corr},0}=\frac{1}{4}$. Because all four states $|\widetilde{\psi}_{bc,0}\rangle$ are equal to $\ket{0,0,0,0}$ when $N=0$, the optimal strategy for discrimination in this case is to randomly guess one of the four states, with an average probability of $\frac{1}{4}$ to be correct. The probability of obtaining a correct result in the subspaces with $N>0$ is equal to $\tilde{P}_{\text{corr},N}=1$. 
There is therefore again a close correspondence between the minimum-error measurement and the optimal unambiguous measurement,
%That is, unless the state is found to contain zero photons, we can perfectly distinguish between the four states. This means that the minimum-error measurement is closely related to an unambiguous measurement. 
similar to what we found for unambiguously distinguishing between the two-mode states $\ket{\alpha, \alpha}$ and $\ket{\alpha,-\alpha}$ by interfering the two modes on a 50/50 beam splitter. There too, unless zero photons are detected at the output, we perfectly distinguish the two states from each other.

\begin{figure}[h]
\centering
\includegraphics[scale=0.5]{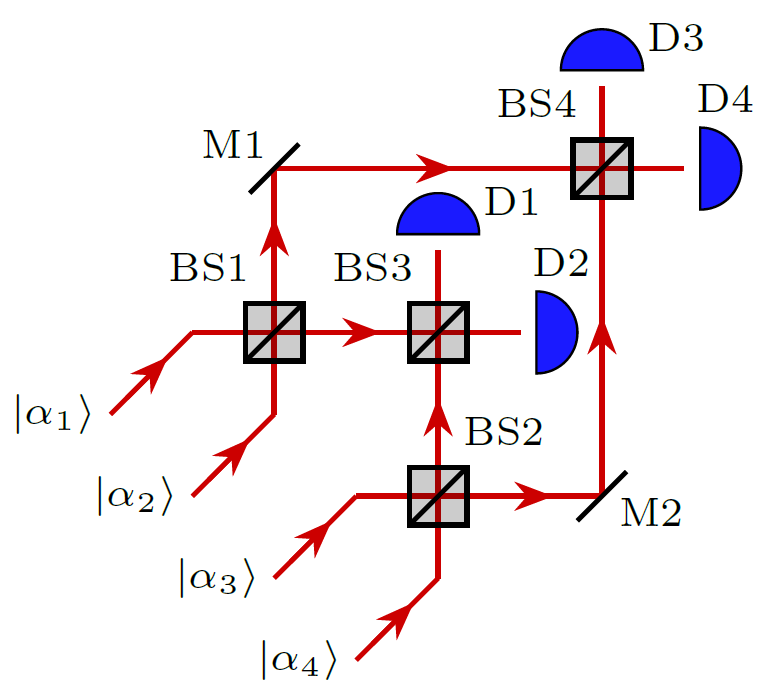}
\caption{Linear-optical setup for distinguishing between the four-mode phase-randomized states in \cref{eq:fourmodemixed}. A click in detectors D1, D2, D3 or D4 uniquely identifies one of the four input states. In the ideal case, it is not possible for more than one detector to click.}
\label{fig3}
\end{figure}

This suggests that there is a linear-optical setup for both the minimum-error and the unambiguous measurement for the four-mode phase-randomized states $\tilde\rho_{bc}$. The setup %for distinguishing between the four-mode phase-randomized states 
is shown in \cref{fig3}. The same setup realises the optimal minimum-error measurement and the optimal unambiguous measurement. The only difference is that detecting no photons is a ``failure" outcome for an unambiguous measurement, while for a minimum-error measurement, we make a random guess. The same setup also realises the optimal unambiguous measurement for the pure states before phase randomization (but not the minimum-error measurement for the pure states). Since there is a phase difference in exactly one of the input mode pairs (1,2) or (3,4), either the top output mode of BS1 and the right output mode of BS2 contain the vacuum state, or vice versa. This means that all of the light exiting beam splitters BS1 and BS2 is either interfered on BS3, with BS4 receiving no photons, or vice versa. For example, the input state
$|\alpha, \alpha, \alpha, -\alpha\rangle$ is transformed to $|\sqrt 2\alpha, 0, 0, \sqrt 2\alpha\rangle$ by beam splitters BS1 and BS2. This state is further changed to $|2\alpha, 0,0,0\rangle$ by BS4; all the light exits into D3, and BS3 receives no light. By checking how the other three input states are transformed, it can be confirmed that a click in D3 uniquely identifies the input state $|\alpha, \alpha, \alpha, -\alpha\rangle$. A click in D4 uniquely identifies $|\alpha, \alpha, -\alpha, \alpha\rangle$, a click in D1 identifies $|\alpha, -\alpha, \alpha, \alpha\rangle$, and a click in D2 identifies $|-\alpha, \alpha, \alpha, \alpha\rangle$.
Just as for the two-mode states, phase-randomization will not affect the number of photons detected at each output.

\begin{figure}[h]
\centering
\includegraphics[scale=0.7]{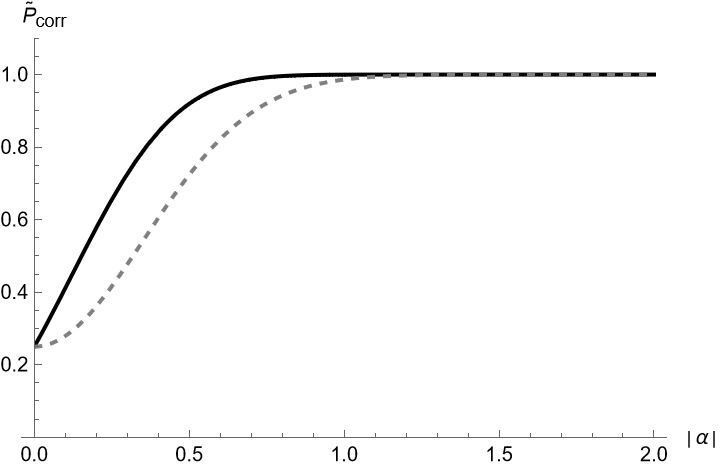}
\caption{Probability of obtaining a correct result $\tilde{P}_{\text{corr}}$ as a function of $\abs{\alpha}$ for the minimum-error measurements for pure (solid black line) and mixed (dashed grey line) four-mode coherent states.}
\label{fig4}
\end{figure}

The probability of obtaining an unambiguous outcome for the four-mode phase-randomized coherent states equals
\begin{equation}
\label{eq:fourmodeprobunambiguous}
\tilde{P}_{\text{s}}=1-e^{-4\abs{\alpha}^2}.
\end{equation}
This is also the success probability for the optimal unambiguous measurement.
The minimum-error probability for correctly discriminating between the four-mode phase-randomized coherent states $\tilde{\rho}_{ab}$ equals
\begin{equation} \label{eq:probcorrfourmode}
\tilde{P}_{\text{corr}}=1-\frac{3e^{-4\abs{\alpha}^2}}{4}.
\end{equation}
The minimum-error measurement for the pure four-mode states is a square-root measurement. The resulting minimum-error probability to correctly distinguish between the respective pure four-mode coherent states is given by (see Appendix \ref{app:protsuccheatprob})
\begin{equation}
\label{eq:probcorrpurefourmode}
\tilde{P}_{\text{corr}}^{\text{pure}}=\frac{1}{16}\big(\sqrt{1+3e^{-4\abs{\alpha}^2}}+3\sqrt{1-e^{-4\abs{\alpha}^2}}\big)^2.
\end{equation}
We plot the probabilities in \cref{eq:probcorrfourmode,eq:probcorrpurefourmode} in \cref{fig4}. As expected, both probabilities start from $1/4$ for $|\alpha|=0$, when the four states are indistinguishable, and increases asymptotically to 1 for $|\alpha|\rightarrow \infty$.

In analogy with the four-mode coherent states, an encoding of two bits using four-dimensional ququart (or two-qubit) states is
%\begin{equation}
%\label{eq:ququartstates}
%\begin{split}
%\ket{\psi_{00}^\rm quq}&=\frac{1}{2}(\ket 0+ \ket 1 +\ket 2 -\ket 3), \\
%\ket{\psi_{01}^\rm quq}&=\frac{1}{2}(\ket 0+ \ket 1 -\ket 2 +\ket 3), \\
%\ket{\psi_{11}^\rm quq}&=\frac{1}{2}(\ket 0- \ket 1 +\ket 2 +\ket 3), \\
%\ket{\psi_{10}^\rm quq}&=\frac{1}{2}(-\ket 0 +\ket 1 +\ket 2 +\ket 3).
%\end{split}
%\end{equation}
\begin{equation}
\label{eq:ququartstates}
\begin{split}
\ket{\psi_{00}^\text{quq}} &= \frac{1}{2}(\ket{0}+ \ket{1} +\ket{2} -\ket{3}), \\
\ket{\psi_{01}^\text{quq}} &= \frac{1}{2}(\ket{0}+ \ket{1} -\ket{2} +\ket{3}), \\
\ket{\psi_{11}^\text{quq}} &= \frac{1}{2}(\ket{0}- \ket{1} +\ket{2} +\ket{3}), \\
\ket{\psi_{10}^\text{quq}} &= \frac{1}{2}(-\ket{0} +\ket{1} +\ket{2} +\ket{3}).
\end{split}
\end{equation}
These states are mutually orthogonal. Because these states therefore are perfectly distinguishable, it is possible to unambiguously determine which single state was received, meaning that it is possible for the receiver to
perfectly learn both bits $x_0$ and $x_1$ of the state $|\psi_{x_0x_1}^{\rm quq}\rangle$. The linear optical setup which achieves this %for the unambiguous discrimination of the four states in \cref{eq:ququartstates} 
is the same as the one used for the four-mode coherent states. The fact that the states in \cref{eq:ququartstates} are orthonormal is connected with the fact that the four-mode coherent states can be unambiguously distinguished from each other as soon as we see a single photon. The qutrit states in \cref{eq:qutritstates}, on the other hand, are non-orthogonal, and it is only possible to unambiguously exclude two of the four states. The linear optical setup that achieves this~\cite{Stroh_2023} does not realize either the minimum-error or the unambiguous measurement for the corresponding phase-randomised three-mode states.

\subsection{Phase-encoded two-mode coherent states}
\label{num4}
\begin{figure}[h]
\centering
\includegraphics[scale=0.7]{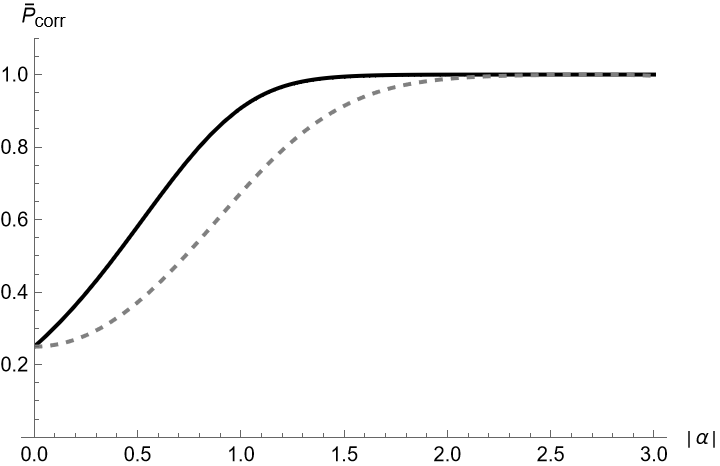}
\caption{Probability of obtaining a correct result $\bar{P}_{\text{corr}}$ as a function of $\abs{\alpha}$ for the four pure (solid black line) and mixed (dashed grey line) phase-encoded coherent states.}
\label{fig5}
\end{figure}

Finally, we examine phase-randomized versions of the four coherent states $|\bar{\psi}_{00}\rangle=\ket{\alpha,\alpha}$, $|\bar{\psi}_{01}\rangle=\ket{\alpha,i\alpha}$, $|\bar{\psi}_{11}\rangle=\ket{\alpha,-\alpha}$ and $|\bar{\psi}_{10}\rangle=\ket{\alpha,-i\alpha}$. 
Either before or after phase-randomization, the phase difference between the modes could encode two classical bits.
Evidently, these states are symmetric, with a unitary symmetry operation that amounts to a phase shift of $i$ on the second mode. This symmetry operation does not change the number of photons in a mode, and can in terms of photon-number states be written 
\begin{equation} \label{eq:phaseencodedunitary}
\bar{U}=\sum_{j,k=0}^\infty i^k\ket{j,k}\bra{j,k}.
\end{equation}
Phase-randomization of the states $|\bar\psi_{bc}\rangle$ gives
\begin{align} \label{eq:phaseencodedmixed}
\bar{\rho}_{bc}&=\frac{1}{2\pi}\int_0^{2\pi}d\theta|\bar{\psi}_{bc}\rangle\langle\bar{\psi}_{bc}| \notag \\
&=e^{-2\abs{\alpha}^2}\sum_{\substack{j,k,p,q=0\\j+k=p+q}}^\infty{\frac{i^{Q_{bc}k}(-i)^{Q_{bc}q}\abs{\alpha}^{j+k+p+q}}{\sqrt{j!k!p!q!}}\ket{j,k}\bra{p,q}}, 
\end{align}
where 
\begin{equation} \label{eq:phaseencodedindex}
Q_{bc}=2bc+\bar{b}c+3b\bar{c}.
\end{equation}
The phase-randomized states $\bar\rho_{bc}$ can again be written as mixtures of pure states $|\bar{\psi}_{bc,N}\rangle$, each one with a different total number of photons.
In the $N$-photon subspace, we have
\begin{equation} \label{eq:phaseencodedNpure}
|\bar{\psi}_{bc,N}\rangle=\frac{1}{\sqrt{\bar{M}_N}}\sum_{\substack{j,k=0\\ j+k=N}}^{N}{\frac{i^{Q_{bc}k}}{\sqrt{j!k!}}\ket{j,k}},
\end{equation}
where $\bar{M}_N=\frac{2^N}{N!}$. We will again obtain the minimum-error probability using the Gram matrix. The total probability for correctly distinguishing between these mixed phase-encoded coherent states is equal to (see Appendix \ref{app:protsuccheatprob})
\begin{alignat}{3}
\label{eq:probcorrphaseencoded}
\bar{P}_{\text{corr}}&=\frac{e^{-2\abs{\alpha}^2}}{4}+\frac{e^{-2\abs{\alpha}^2}}{8}&& \notag \\
&\times\sum_{N=1}^\infty\frac{(2\abs{\alpha}^2)^N}{N!}\bigg[\sqrt{1+\sqrt{1-2^{-N+2}\cos^2{(N\pi/4)}}}&& \notag \\
& +\sqrt{1+\sqrt{1-2^{-N+2}\sin^2{(N\pi/4)}}}\bigg]^2.&&
\end{alignat}
The probability to correctly distinguish between the corresponding pure phase-encoded two-mode coherent states is given by (see Appendix \ref{app:protsuccheatprob})
\begin{equation}
\label{eq:probcorrpurephaseencoded}
\begin{split}
\bar{P}_{\text{corr}}^{\text{pure}}=&\frac{e^{-\abs{\alpha}^2}}{4}\bigg(\sqrt{\cosh{\abs{\alpha}^2}+\sqrt{\cosh^2{\abs{\alpha}^2}-\cos^2{\abs{\alpha}^2}}}\\
&+\sqrt{\sinh{\abs{\alpha}^2}+\sqrt{\sinh^2{\abs{\alpha}^2}-\sin^2{\abs{\alpha}^2}}}\ \bigg)^2.
\end{split}
\end{equation}
We plot \cref{eq:probcorrphaseencoded,eq:probcorrpurephaseencoded} as a function of $|\alpha|$ in \cref{fig5}. Again, we can see how phase-randomization lowers the minimum error probability.

\begin{figure}[h]
\centering
\includegraphics[scale=0.7]{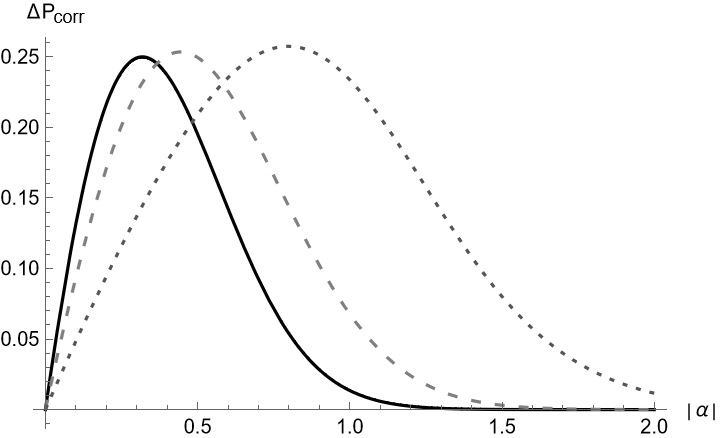}
\caption{Differences between the probabilities to be correct $\Delta P_{\text{corr}}$ for pure and phase-randomized three-mode (solid black line), four-mode (medium dashed grey line) and phase-encoded (small dashed dark grey line) coherent states. The differences reach a maximum value of about 0.25 
%are maximal around $\frac{1}{4}$ 
for all cases, but this maximum occurs for different values of $\abs{\alpha}$.}
\label{fig6}
\end{figure}

To quantify the effect of mixing, we define the difference between the probabilities to be correct for the pure and the mixed cases, for each set of quantum states, as $\Delta P_{\text{corr}}=P_{\text{corr}}^{\text{pure}}-P_{\text{corr}}$. We plot $\Delta P_{\text{corr}}$ for the three-mode, the four-mode and the phase-encoded coherent states that we examined in \cref{fig6}.

\section{Application to quantum oblivious transfer and quantum retrieval games}
\label{sec:OT}

\subsection{Quantum oblivious transfer}

Distinguishing between four (often symmetric) states is relevant for quantum 1-out-of-2 oblivious transfer~\cite{Stroh_2023,OT2021}. This is a cryptographic primitive where a sender has two (classical) bits, and a receiver should receive one of them, but obtain no information about the other bit. The sender Alice should not know which bit the receiver Bob has obtained. The interest in oblivious transfer stems from the fact that it is a universal building block for multi-party computation, where two or more mistrustful parties want to compute something together, without revealing any more than necessary about their individual input data. Oblivious transfer is possible  ``classically" (without the use of quantum resources) with computational security. If one desires information-theoretic security, % for an oblivious transfer protocol, 
then only trivial ``classical" protocols where either the sender or the receiver can cheat perfectly are possible (if we restrict to protocols where the receiver always correctly obtains a bit value if both parties are honest). If the classical bit values are encoded in quantum states, then perfect oblivious transfer still is not possible, unless one restricts the quality or amount of quantum memory the parties have access to. Imperfect quantum oblivious transfer is however still possible, where neither the sender nor receiver can cheat perfectly~\cite{reichmuth2024incompletequantumoblivioustransfer,10.5555/2481591.2481600,Chailloux2013OptimalBF,OT2021,Kundu2022deviceindependent,Stroh_2023,PhysRevResearch.6.043004,unknown}. They might still cheat, but their probability to do so is limited.

Until now, the quantum states which have been considered and analyzed for quantum oblivious transfer were almost always pure and symmetric. Mixed states can however sometimes lead to improved performance~\cite{reichmuth2024incompletequantumoblivioustransfer}. Hence it is of interest to examine whether phase-randomized multi-mode coherent states could be of use.
The sender's bit values would be encoded in four quantum states. 
In 1-out-of-2 oblivious transfer, an honest receiver should only learn one of the sender's bit values, whereas a dishonest receiver wants to learn both bit values. 
The minimum-error probability for optimally distinguishing between these four quantum states, which we above denoted by $P_{\text{corr}}$, then corresponds to the cheating probability for the receiver. %$P_{\text{corr}}$ then gives the cheating probability $B_{\text{OT}}$. 
To assess how suitable a particular set of states is for quantum oblivious transfer, we also need the probability for an honest receiver to correctly learn (only) one bit value, and the probability for the sender to correctly guess which bit the receiver has received.

Often an honest party actively chooses which bit they want to learn, the first or the second. This is attractive especially in non-interactive quantum oblivious transfer protocols, where a quantum state is sent from one party to the other and then measured, with no further communication taking place, because if the receiving party randomly and actively chooses between two measurements, the sending party can obtain no information about the receiver's choice of measurement. The price one has to pay is that even if both sender and receiver are honest, the probability for the receiver to obtain reliable information about a bit they actively choose is generally lower than if they randomly obtain information about either bit value~\cite{Amiri_2017, Stroh_2023}.

The receiver's optimal measurement for learning one of the bits, with an active choice of which bit the receiver wishes to learn, is a Helstrom measurement.
Their probability $P_{\text{1bit}}$ to obtain this bit value correctly then corresponds to the probability that the protocol works as intended if both parties are honest. %(sometimes referred to as ``completeness" of the protocol). 
For example, for an oblivious transfer protocol using pure symmetric states, to learn the first bit, an honest Bob makes the Helstrom measurement to distinguish between the two equiprobable mixed states 
\begin{equation}
    \begin{split}
\rho_0&=\frac{1}{2}(\ket{\psi_{00}}\bra{\psi_{00}}+\ket{\psi_{01}}\bra{\psi_{01}})\\
\rho_1&=\frac{1}{2}(\ket{\psi_{10}}\bra{\psi_{10}}+\ket{\psi_{11}}\bra{\psi_{11}}),
\end{split}
\end{equation}
where $\ket{\psi_{x_0 x_1}}$ is the state encoding two bits $x_0$ and $x_1$.
%specifically for the first bit here, which corresponds to the success of the protocol ($P_{\text{s}}^{\text{1bit}}\equiv P_{\text{s}}$).\\
%Using the notation $\ket{\psi_K}=U^K\ket{\psi_0}$ with $K=0,1,2,3$ for the four states of the protocol, where in two-bit labeling we have $\ket{\psi_0}\equiv\ket{\psi_{00}}$, $\ket{\psi_1}\equiv\ket{\psi_{01}}$, $\ket{\psi_2}\equiv\ket{\psi_{11}}$ and $\ket{\psi_3}\equiv\ket{\psi_{10}}$, we present the general structure of the Gram matrix 
%\begin{equation}
%\label{eq:generalgrammatrix}
%\mathcal{G}=\begin{pmatrix}
%\bra{\psi_0}\ket{\psi_0} & \bra{\psi_0}\ket{\psi_1} & %\bra{\psi_0}\ket{\psi_2} & \bra{\psi_0}\ket{\psi_3} \\
%\bra{\psi_1}\ket{\psi_0} & \bra{\psi_1}\ket{\psi_1} & %\bra{\psi_1}\ket{\psi_2} & \bra{\psi_1}\ket{\psi_3} \\
%\bra{\psi_2}\ket{\psi_0} & \bra{\psi_2}\ket{\psi_1} & \bra{\psi_2}\ket{\psi_2} & \bra{\psi_2}\ket{\psi_3} \\
%\bra{\psi_3}\ket{\psi_0} & \bra{\psi_3}\ket{\psi_1} & \bra{\psi_3}\ket{\psi_2} & \bra{\psi_3}\ket{\psi_3} \\
%\end{pmatrix}=\begin{pmatrix}
%1 & F & G & F^* \\
%F^* & 1 & F & G \\
%G & F^* & 1 & F \\
%F & G & F^* & 1
%\end{pmatrix}
%\end{equation}
For a protocol that uses pure symmetric states, where the sender chooses $x_0$ and $x_1$ uniformly at random, this probability is~\cite{reichmuth2024incompletequantumoblivioustransfer} (see also Appendix \ref{app:protsuccheatprob})
%\begin{equation} \label{eq:protocolsucprob}
%\begin{split}
%P_{\text{1bit}}&=\frac{1}{2}\bigg[1+\frac{1}{2}\sqrt{1-G^2+2\sqrt{(1+G)^2(\text{Im}F)^2+(1-G)^2(\text{Re}F)^2-4(\text{Re}F)^2(\text{Im}F)^2}} \\
%&+\frac{1}{2}\sqrt{1-G^2-2\sqrt{(1+G)^2(\text{Im}F)^2+(1-G)^2(\text{Re}F)^2-4(\text{Re}F)^2(\text{Im}F)^2}}\bigg]. 
%\end{split}
%\end{equation}
\begin{widetext}
\begin{equation}
\label{eq:protocolsucprob}
\begin{split}
2(2P_{\text{1bit}}-1)=
&\sqrt{1-G^2+\sqrt{(1+G)^2(\text{Im}F)^2+(1-G)^2(\text{Re}F)^2 -4(\text{Re}F)^2(\text{Im}F)^2}} \\
&+\sqrt{1-G^2-\sqrt{(1+G)^2(\text{Im}F)^2+(1-G)^2(\text{Re}F)^2-4(\text{Re}F)^2(\text{Im}F)^2}}
\end{split}
\end{equation}
Here $F=\bra{\psi_{00}}\ket{\psi_{01}}$, $G=\bra{\psi_{00}}\ket{\psi_{11}}$ and $F^*=\bra{\psi_{00}}\ket{\psi_{10}}$ are the pairwise overlaps between different states.
\end{widetext}
%\Ioannis{\begin{equation}
%\label{eq:protocolsucprob}
%\begin{split}
%2&(2P_{\text{1bit}}-1)=\\ &\sqrt{1-G^2+\sqrt{\begin{aligned}(1+G)^2(\text{Im}F)^2&+(1-G)^2(\text{Re}F)^2 \\ &-4(\text{Re}F)^2(\text{Im}F)^2\end{aligned}}} \\
%+&\sqrt{1-G^2-\sqrt{\begin{aligned}(1+G)^2(\text{Im}F)^2&+(1-G)^2(\text{Re}F)^2 \\ &-4(\text{Re}F)^2(\text{Im}F)^2\end{aligned}}}
%\end{split}
%\end{equation}}
%\Ioannis{ \bf 19 August: I think we should use the single column format for the equation of $P_{\text{1bit}}$. \Erika{Split lines inside the square roots is a little unconventional and might get changed by the journal (or might not).}} \Ioannis{ \bf 21 August: I meant keeping the other format. The single column format which you wrote alternatively.} 

%As shown in~\cite{OneSided2021} \Ioannis{\bf This one-sided paper is not even in arxiv.}, for a protocol that uses pure symmetric states, the success probability of the protocol in terms of the pairwise overlaps $F$ and $G$ of the Gram matrix is given by (see Appendix \ref{app:protsuccheatprob})

To find an honest Bob's probability to obtain one bit for the different sets of phase-randomized (and thus mixed) symmetric multi-mode coherent states in the previous section, we note that the Helstrom measurement for distinguishing between the mixed states corresponding to the two values for the bit Bob wishes to learn can be realized by first measuring the total photon number, and then distinguishing between the resulting possible states in the relevant photon-number subspace.
This is the optimal measurement, because the relevant density operators are block-diagonal with respect to total photon number. That is, because the phase-randomized states can be written as $\rho_{b}=\sum_Np_N\rho_{b,N}$, where $b$ is the value of the bit Bob wishes to learn, and $\rho_{b,N}$ is a state with exactly $N$ photons.  For example, if an honest Bob wishes to find $x_0$, %$P_{\textrm{1bit}}$,
he makes the Helstrom measurement (assuming he measured $N$ photons) for distinguishing between the equiprobable mixed states 
\begin{equation}
\begin{split}
\rho_{0,N}&=\frac{1}{2}(\ket{\psi_{00,N}}\bra{\psi_{00,N}}+\ket{\psi_{01,N}}\bra{\psi_{01,N}})\\
\rho_{1,N}&=\frac{1}{2}(\ket{\psi_{10,N}}\bra{\psi_{10,N}}+\ket{\psi_{11,N}}\bra{\psi_{11,N}})
\end{split}
\end{equation}
with the probability $P_{\textrm{1bit},N}$ of giving a correct result. Then, in total, Bob's probability to correctly learn the bit value of his choice is
\begin{equation}
P_{\textrm{1bit}}=\sum_{N=0}^\infty p_NP_{\textrm{1bit},N}.
\end{equation}
Therefore, to find the probability for Bob to correctly learn one bit of his choice, we first find his probability to correctly distinguish between the two relevant mixed states
in each $N$-photon subspace. %, that is, between the two states in each subspace, corresponding to the different values of the bit Bob wishes to learn. 
Then, we sum these probabilities, multiplied with the appropriate probabilities for each photon-number subspace, in analogy with how we obtained the optimal minimum-error probability for distinguishing between all four phase-randomized states in the previous section.

\begin{figure}[h]
\centering
\includegraphics[scale=0.7]{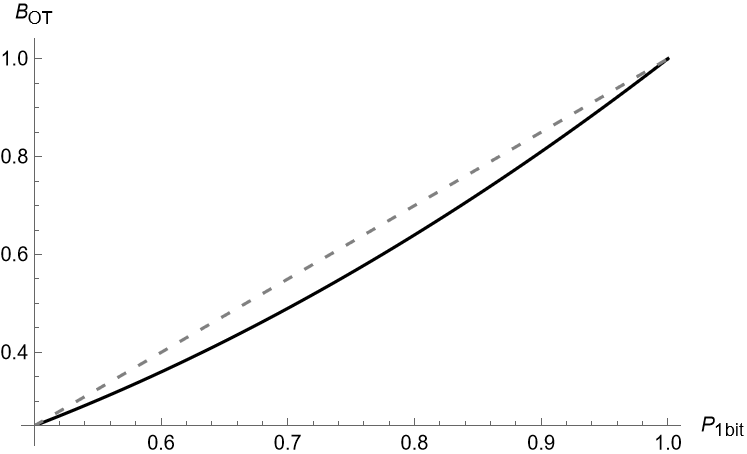}
\caption{Cheating probability $B_{\text{OT}}$ as a function of protocol success probability $P_{\text{1bit}}$, for pure (solid black line) and phase-randomized (dashed grey line) three-mode coherent states. We observe that employing phase randomization 
for the three-mode coherent states leads to a higher cheating probability as compared to using pure states.}
\label{fig7}
\end{figure}

For the three-mode phase-randomized coherent states in \cref{eq:threemodemixed}, we have $F_N=(\frac{1}{3})^N$ and $G_N=(-\frac{1}{3})^N$. The probability for Bob to correctly obtain one bit value, for the $N$-photon subspace, is given by 
%\Erika{\bf Are the two resulting states in each subspace mixed? (I think I added ``mixed" above when editing just now, but can't quite remember.) They should be mixed, because Bob wants only one bit. That is, is the expression below what one obtains for a Helstrom measurement in each photon-number subspace? If yes, we might be clearer about it. It almost sounds like we have pure states in each subspace (since we are talking about $F_N$ and $G_N$, which are pure-state overlaps).}
\begin{equation}
\begin{split} \label{eq:protocolsucprobthreemodeN}
P_{\text{1bit},N}=\frac{1}{2}\bigg[1+\frac{1}{2}&\Big[\sqrt{G_N^2-2G_N+1} \\ &+\sqrt{-3G_N^2+2G_N+1}\ \Big]\bigg].
\end{split}
\end{equation}
Bob's overall probability to obtain one correct bit value for the three-mode coherent states is then
\begin{alignat}{3}
\label{eq:protocolsucprobthreemode}
P_{\text{1bit}} & =\sum_{N=0}^\infty p_NP_{\text{1bit},N} \notag \\ & =\frac{1}{2}\ +\ \frac{e^{-3\abs{\alpha}^2}}{4}\sum_{N=0}^\infty&&\frac{(3\abs{\alpha}^2)^N}{N!}\Big[\sqrt{G_N^2-2G_N+1} \notag \\ & &&+\sqrt{-3G_N^2+2G_N+1}\ \Big].
\end{alignat}
For the four-mode phase-randomized coherent states in \eqref{eq:fourmodemixed}, whenever nonzero photons are detected, the state can be uniquely identified. That is, it is no more difficult to obtain both bit values than one. Hence we do not expect this set of states to be of interest for quantum oblivious transfer. Nevertheless, the probability to correctly guess one (or both) bits %The protocol success probability \Ioannis{for the phase-randomized four-mode states} 
would be (see Appendix \ref{app:protsuccheatprob})
\begin{equation} 
\label{eq:protocolsucprobfourmode}
\tilde{P}_{\text{1bit}}=1-\frac{e^{-4\abs{\alpha}^2}}{2}.
\end{equation}
For the phase-randomized phase-encoded states in \cref{eq:phaseencodedmixed}, the probability for Bob to obtain one bit correctly is (see Appendix \ref{app:protsuccheatprob})
\begin{align} \label{eq:protocolsucprobphaseencoded}
\bar{P}_{\text{1bit}}&=\frac{1}{2}+\frac{e^{-2\abs{\alpha}^2}}{2\sqrt{2}}\sum_{N=1}^\infty\frac{(2\abs{\alpha}^2)^N}{N!}\times \notag \\ &\times\sqrt{1+\sqrt{4^{-N+1}\sin^2{(N\pi/2)}-2^{-N+2}+1}}.
\end{align}
To evaluate how well a specific set of for states works for realizing quantum oblivious transfer, we find the relation between $B_{\text{OT}}$ and $P_{\text{1bit}}$ in each case. In particular, since it is known that mixed states can sometimes outperform pure states, we are interested in whether phase-randomisation can improve protocol performance. Also, it is interesting to know whether a protocol still works acceptably well also if the phase is randomized, since this would imply robustness against noise. The lower the receiver's cheating probability is, for a fixed success probability for an honest receiver, the better the protocol is. For the three-mode phase-randomized states, we find the relation
\begin{equation} \label{eq:threemodemixedcheatprobvsprotocolsucprob}
B_{\text{OT}}=\frac{3}{2}P_{\text{1bit}}-\frac{1}{2},
\end{equation}
%In the pure case scenario, 
while for the corresponding three-mode pure states we have 
\begin{equation}
B_{\text{OT}}^{\text{pure}}=(P_{\text{1bit}}^{\text{pure}})^2<\frac{3}{2}P_{\text{1bit}}^{\text{pure}}-\frac{1}{2}
\end{equation}
for $\frac{1}{2}<P_{\text{1bit}}^{\text{pure}}<1$. Therefore, phase-randomization makes the protocol performance worse for this particular set of states (\cref{fig7}). However, protocol performance is not worsened as much as one might have expected.

For the four-mode states, the relation for both pure and mixed states is the same as in \cref{eq:threemodemixedcheatprobvsprotocolsucprob}, %\Erika{\bf Is this equation reference correct? It seems it should be \ref{eq:protocolsucprobfourmode}?} \Ioannis{ \bf 21 August: I think it's correct. The relation between $B_\text{OT}$ and $P_\text{1bit}$ for both pure and mixed states is $B_\text{OT}=\frac{3}{2}P_\text{1bit}-\frac{1}{2}$ which is equation $(50)$.} so mixing these states makes no difference to the protocol (in either case, the receiver can cheat perfectly, as we already implied).
%\Erika{\bf Therefore, should we not mention this example here anymore?} \Ioannis{\bf We can say that for the four-mode case when mixing makes no difference, in that case, we can still get an advantage from the fact that we don't need the phase reference. \Erika{Hardly an ``advantage", since the protocol does not work at all (receiver cheats perfectly) either with pure or mixed states. It is not possible to worsen performance with mixing, since the performance is already as bad as it can be...}} \Ioannis{ \bf 19 August: If you think we should erase the four-mode comment here, let's do it. I included this comment with bold red only because you had mentioned it in one previous meeting.}

\begin{figure}[h]
\centering
\includegraphics[scale=0.7]{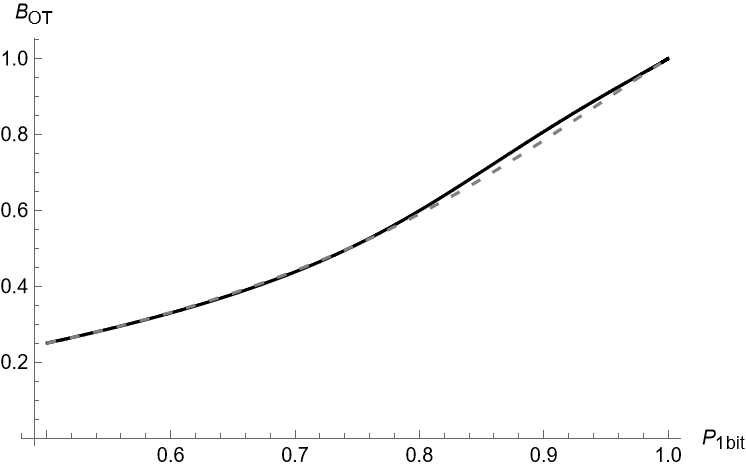}
\caption{Cheating probability $B_{\text{OT}}$ as a function of  protocol success probability $P_{\text{1bit}}$, for pure (solid black line) and phase-randomized (dashed grey line) two-mode phase-encoded coherent states. We notice that mixedness marginally lowers the cheating probability $B_{\text{OT}}$ for $P_{\text{1bit}}\gtrsim 0.8$.}
\label{fig8}
\end{figure}

For the two-mode phase-encoded coherent states in \eqref{eq:phaseencodedmixed}, because it is difficult to find an analytical form for the relation between $B_{\text{OT}}$ and $P_{\text{1bit}}$, we plot $B_{\text{OT}}$ against $P_{\text{1bit}}$. We note that mixedness helps marginally in some regions, but makes protocol performance worse in others (\cref{fig8}). A protocol is better if, for the same probability $P_{\text{1bit}}$ for an honest receiver Bob to obtain one bit value correctly, the cheating probability for a dishonest receiver to correctly guess both bit values is as low as possible. The results are explained in more detail in Appendix \ref{app:protsuccheatprob}.


\subsection{Quantum retrieval games}
\label{num5}

The measurements we have been investigating may also be relevant for so-called quantum retrieval games~\cite{DBLP:conf/coco/Gavinsky12, Arrazola_2016}. %\Erika{\bf Is this the original reference or a later one?} \Ioannis{\bf 19 August: Yes, I think $[35]$ is the original reference.} 
In a quantum retrieval game, one party, whom we will here call sender (Alice), encodes a number of classical bits in a quantum state. The quantum states are selected from a set of non-orthogonal states. The state is sent to another party, whom we will call receiver (Bob). The receiver selects a measurement, among some possible measurements, so that the measurement will give information about some of the sender's bit values, or some of their properties, but not about others. 
%Usually there is a probabilistic element to what the sender might learn even after selecting the measurement. 
For example, the sender might select four classical bits $x_0, x_1, x_2$ and $x_3$, and encode these in a state
%\begin{equation}
%    \ket{\psi_{x_0, x_1, x_2, x_3}}=\frac{1}{2} \left[(-1)^{x_0}\ket 0 +(-1)^{x_1}\ket 1 +(-1)^{x_2}\ket 2 +(-1)^{x_3}\ket 3\right].
%\end{equation}
\begin{equation}
\label{eq:QRGstates}
\ket{\phi_{x_0x_1x_2x_3}}=\frac{1}{\sqrt{4}}\sum_{i=0}^{3}(-1)^{x_i}\ket{i}.
\end{equation}
The receiver then chooses a matching, pairing up the four bits. 
%The receiver's measurement gives information about the XOR of the bits in one pair. 
In this example, the receiver might choose to pair $x_0$ with $x_1$, and $x_2$ with $x_3$. A measurement in the basis 
$\{\frac{1}{\sqrt 2}(\ket 0 \pm \ket 1), \frac{1}{\sqrt 2}(\ket 2 \pm \ket 3)\}$ would then give the receiver information about either the relative phase of the basis states $\ket 0$ and $\ket 1$ (that is, about $x_0\oplus x_1$), or about the relative phase of the basis states $\ket 2$ and $\ket 3$ (that is, about $x_2\oplus x_3$). A different choice for the receiver would be to pair $x_0$ with $x_2$, and $x_1$ with $x_3$. The last option is to pair $x_0$ with $x_3$, and $x_1$ with $x_2$. Because the 16 possible states are non-orthogonal, the receiver cannot learn all four bits.

A quantum retrieval game is reminiscent of 1-out-of-$n$ XOR oblivious transfer in the sense that the sender does not know exactly what the receiver learns, and the receiver cannot perfectly learn all of the sender's bit values~\cite{Lara_thesis}. The former property is usually not used or even mentioned e.g. in schemes for quantum money~\cite{DBLP:conf/coco/Gavinsky12,doi:10.1073/pnas.1203552109,georgiou_et_al:LIPIcs.TQC.2015.92,Amiri_2017,Guan_2018}, which can be constructed using quantum retrieval games. If the receiver cheats and tries to correctly guess all of the sender's bit values, then their optimal measurement is a minimum-error measurement. Since retrieval games often use symmetric states, the measurements we have derived may be relevant.

%For the cryptographic primitive of quantum retrieval games (QRGs), 
A possible set of states to use for a retrieval game might be the qutrit states in Eq. \eqref{eq:qutritstates}. % and the ququart \eqref{eq:ququartstates} states which we analyzed. 
 In hidden matching quantum retrieval games~\cite{Arrazola_2016}, the sender Alice usually picks $n$-bit strings where $n$ is even, meaning that there are $2^n$ different states, and that all bits can be paired up. % so that the matching is perfect. Therefore, 
A retrieval game based on our qutrit states would be somewhat different, % from the usual definition, 
in that the receiver Bob does not choose a pairing, but instead probabilistically obtains either the first bit, the second bit, or their XOR~\cite{Stroh_2023}. We could also include four more states of the same form of the qutrit states, but with $-\ket 0$ instead of $\ket 0$, and consider that the sender has chosen three bits instead of two. The receiver would then probabilistically obtain information about the XOR either of the first and the second bits, or of the second and third bits, or the first and the third bits.

As for the four ququart states in Eq. \eqref{eq:ququartstates}, they are a subset of the total set of 16 states in hidden matching quantum retrieval games with $4$-bit strings.
%In this type of hidden matching retrieval games, Alice has inputs with equal probability from the set of all possible $4$-bit strings, i.e. $X=\{0,1\}^4$. Then, she encodes her inputs in the pure states
%\begin{equation}
%\label{eq:QRGstates}
%\ket{\phi_{x_0x_1x_2x_3}}=\frac{1}{\sqrt{4}}\sum_{i=0}^{3}(-1)^{x_i}\ket{i}
%\end{equation}
%where $x_i$ is the $i$-th bit value of the string $x$ and she prepares each state with probability $\frac{1}{2^4}$. 
%The concept of hidden matching QRGs is quite similar to the one of $1$-out-of-$n$ XOT~\cite{Lara_thesis} and is widely used for quantum money schemes~\cite{Amiri_2017,Guan_2018}. In a $4$D Hilbert space, perfect distinction among 16 states isn't possible. While these states can encode up to 4 classical bits, only 2 bits can be retrieved (Holevo bound) due to their nonorthogonality. 
The four ququart states in Eq. \eqref{eq:ququartstates} can be perfectly distinguished, meaning that the two classical bits one could encode in these four states can both be perfectly retrieved by the receiver. %If only these four states were to be used, Bob could retrieve both bits, obtaining all the information. These 4 states cannot be used in the quantum retrieval game as they're a subset of the total 16 states, where the objective is partial retrieval. 
Perfect knowledge allows unlimited copying of the state, which is undesirable in schemes for quantum money or tokens, since one wants to prevent copying. In order to construct a viable quantum money or token scheme, we would need to add more states to the set. If the resulting set of states is symmetric, then similar techniques as we have employed in this paper could be used to obtain optimal measurements especially for a  cheating receiver, who wants to distinguish between all possible states.

\section{Discussion and future research}

We have investigated optimal quantum measurements for distinguishing between phase-randomized multi-mode coherent states.
Phase-randomized weak coherent states are used in quantum communication, especially in quantum key distribution (QKD). %, serving as a cornerstone for ensuring communication security. 
They have been demonstrated to significantly 
enhance the performance of QKD protocols%, as highlighted in seminal works such as
~\cite{lo2007security}. In quantum key distribution, security initially relied on the use of single-photon sources. However, perfect single-photon sources are difficult to realize.
As an alternative, weak lasers are commonly employed, and can be modeled as coherent states with random 
phases. The security of QKD protocols depends on the assumption that these phases are uniformly random~\cite{lo2007security}.
However, achieving continuous random phase poses challenges, and imperfect phase-randomization significantly 
undermines the security of QKD~\cite{lo2007security}. Several methods have been explored to address this issue, with one prominent approach being the consideration of a limited number of discrete phases~\cite{Cao_2015,Yuan2016,Zhao_2007}.

Phase-randomized weak coherent states play an indispensable role also in other QKD techniques, including decoy-state QKD. Without phase randomization, the equations for the gain and the quantum bit error rate of the signal state, which are crucial in the theory of decoy states, cannot be obtained~\cite{Lo_2005}. Additionally, phase-randomized weak coherent states are essential for protocols such as twin-field QKD. Phase-randomization allows Alice and Bob to apply the decoy-state method so that they can detect eavesdropping attempts by an adversary, often referred to as Eve. When Alice and Bob both choose the Z-basis which is for testing potential tampering by Eve, because they use phase-randomized coherent states, they can instead use decoy state intensity settings that they have at their disposal. This is based on the fact that Eve 
cannot distinguish between the signal and decoy states. The more decoy intensities Alice and Bob use, the better they estimate the yields, i.e. the detection statistics, and consequently the phase-error rate of the key, thus bounding the information that could have been leaked to a potential eavesdropper and achieving a higher key rate~\cite{Lucamarini_2018,curty2018simple}. %Without phase randomization, we cannot express these states in the basis complementary to coherent states.

When coherent states are used, a shared reference frame between the two communicating parties typically involves a strong pulse acting as a reference and a weak signal pulse, where the phase relationship between the two pulses is known~\cite{Bennett1992,Townsend1993}. However, the use of phase-randomized coherent states implies scenarios where parties lack a shared frame of reference. Related to this, it has been shown that communication with no shared frame of reference can be achieved with entangled discrete-level states~\cite{Bartlett2003}. 
%In~\cite{Mahdavifar_2021}, it is argued that phase-randomized states can produce entangled states with some additional unwanted separable states, the contribution of which can be effectively removed from the final result. \Erika{\bf Need to check the preceding publication.} \Ioannis{ \bf 19 August: We both agreed that this reference should not be included.} 
Consequently, phase-randomized coherent states offer a promising avenue for facilitating communication in situations where shared reference frames are absent. Another study further highlights the broad applicability of phase-randomized coherent states in communication protocols~\cite{Allevi_2012}. %\Erika{\bf Does this paper talk about ``numerous" studies (maybe like ten or more)? Is it not just one study? Or is it a reviewpaper, which might then mention ``numerous" studies?} \Ioannis{\bf This is not a review paper. }

It is also generally of interest to know how well it is possible to distinguish between mixed quantum states, because noise usually leads to mixed states, and in realistic problems we therefore usually deal with mixed states.
Our method of finding optimal measurements can be applied whenever mixing generates photon states that separate in subspaces with different total photon number $N$. A similar approach can also be applied for non-symmetric mixed states, if they are block diagonal in different photon-number subspaces. Finally, exactly how mixed quantum states may be of use in quantum communication schemes such as oblivious transfer is still an open question. For the cases we examined, we found that perhaps counter-intuitively, mixing did not significantly worsen protocol performance and can even give a slight advantage in some parameter regions.

\acknowledgments
EA would like to acknowledge support by the UK Engineering and Physical Sciences Research Council (EPSRC) under Grant No. EP/T001011/1.

\bibliographystyle{unsrt}
\bibliography{Refs}

\appendix

%\section{Proof that the square-root measurement is optimal for the two-mode states in eq. \eqref{eq:twomixed}}

\section{Optimal measurement for the phase-randomized sets of states}
\label{app:general}

Starting from a general multi-mode coherent state
\begin{equation}
\label{eq:multimodecoherent}
\ket{\psi_0}=\ket{\alpha_1,\alpha_2,\ldots,\alpha_M}
\end{equation}
we express the set of pure symmetric states as
\begin{equation}
\label{eq:symmetricity}
\ket{\psi_K}=U^K\ket{\psi_0},\quad K=0,1,\ldots,L-1
\end{equation}
with $U^L=\mathds{1}$, $\alpha_j=\abs{\alpha_j}e^{i(\theta+\Delta\theta_j)}$ and $\Delta\theta_1=0$. We expand the state as
%\Ioannis
{\begin{align}
\label{eq:multimodesplitinphotonsubspaces}
\ket{\psi_0}&%=\ket{\alpha_1,\alpha_2,\ldots,\alpha_M}
=e^{\frac{-(\abs{\alpha_1}^2+\cdots+\abs{\alpha_M}^2)}{2}}\times \notag \\ &\times\sum_{n_1,n_2,\ldots,n_M=0}^\infty\frac{\alpha_1^{n_1}\alpha_2^{n_2}\cdots \alpha_M^{n_M}}{\sqrt{n_1!n_2!\cdots n_M!}}\ket{n_1,n_2,\ldots,n_M} \notag \\
%&=\bigoplus_{N=0}^\infty \sum_{\substack{n_1,n_2,\ldots,n_M=0\\n_1+n_2+\cdots+n_M=N}}^Ne^{\frac{-(\abs{\alpha_1}^2+\cdots+\abs{\alpha_M}^2)}{2}}\frac{\alpha_1^{n_1}\alpha_2^{n_2}\cdots\alpha_M^{n_M}}{\sqrt{n_1!n_2!\cdots n_M!}}\ket{n_1,n_2,\ldots,n_M}\\
%&=\bigoplus_{N=0}^\infty\sum_{\substack{n_i=0,i\in[M]\\\sum_kn_k=N}}^Ne^{\frac{-(\abs{\alpha_1}^2+\cdots+\abs{\alpha_M}^2)}{2}}\frac{\abs{\alpha_1}^{n_1}\abs{\alpha_2}^{n_2}\cdots \abs{\alpha_M}^{n_M}}{\sqrt{n_1!n_2!\cdots n_M!}}\prod_{m=1}^Me^{i(\theta+\Delta\theta_m)n_m}\ket{n_1,n_2,\ldots,n_M}\\
&=\bigoplus_{N=0}^\infty\sum_{\substack{n_i=0,i\in[M]\\\sum_kn_k=N}}^NA(\vec{r},\vec{n})B(\theta,\vec{n})\ket{n_1,n_2,\ldots,n_M} \notag \\
&=\bigoplus_{N=0}^\infty\sqrt{p_N}\ket{\psi_{0,N}},
\end{align}}
where for $\vec{n}=(n_1,n_2,\ldots,n_M)$ and $\vec{r}=(\abs{\alpha_1},\abs{\alpha_2},\ldots,\abs{\alpha_M})$, we have
%\Ioannis
{\begin{equation}
\label{eq:magnitudeandphaseparts}
\begin{split}
A(\vec{r},\vec{n})&=e^{\frac{-(\abs{\alpha_1}^2+\cdots+\abs{\alpha_M}^2)}{2}}\frac{\abs{\alpha_1}^{n_1}\abs{\alpha_2}^{n_2}\cdots \abs{\alpha_M}^{n_M}}{\sqrt{n_1!n_2!\cdots n_M!}}, \\ B(\theta,\vec{n})&=\prod_{m=1}^Me^{i(\theta+\Delta\theta_m)n_m}.
\end{split}
\end{equation}}
The probability for $N$ photons %distribution 
is given by
%\Ioannis
{\begin{align}
\label{eq:poissondistribution}
p_N&=e^{-(\abs{\alpha_1}^2+\cdots+\abs{\alpha_M}^2)}\frac{(\abs{\alpha_1}^2+\cdots+\abs{\alpha_M}^2)^N}{N!} \notag \\ &=e^{-\langle N\rangle}\frac{{\langle N\rangle}^N}{N!},
\end{align}}
which is a Poisson distribution. The states $\ket{\psi_{0,N}}$ are
\begin{equation}
\label{eq:puremultimodecoherentstatesnsubspace}
\ket{\psi_{0,N}}=\frac{1}{\sqrt{p_N}}\sum_{\substack{n_i=0,i\in[M]\\\sum_kn_k=N}}^NA(\vec{r},\vec{n})B(\theta,\vec{n})\ket{n_1,n_2,\ldots,n_M}.
\end{equation}
If the unitary operator $U$ does not change the total number of photons, i.e.
%\Ioannis
{\begin{equation}
\label{eq:unitaryNsubspacesplit}
\begin{split}
U=\bigoplus_{N=0}^\infty&\sum_{\substack{b_1,\ldots,b_M=0\\c_1,\ldots,c_M=0\\\sum_kb_k=\sum_lc_l=N}}^NC_{b_1,\ldots,b_M}^{c_1,\ldots,c_M}\times \\ &\times\ket{b_1,b_2,\ldots,b_M}\bra{c_1,c_2,\ldots,c_M}=\bigoplus_{N=0}^\infty U_N,
\end{split}
\end{equation}}
then it holds that
\begin{equation}
\label{eq:unitaryoperatorpower}
\begin{split}
%\Rightarrow\\
U^K&=\bigoplus_{N_1,\ldots,N_K=0}^\infty U_{N_1}\cdots U_{N_K}=\bigoplus_{N=0}^\infty (U_N)^K.
\end{split}
\end{equation}
The first phase-randomized state is
\begin{equation}
\label{eq:firstphaserandomizedstate}
\rho_0=\frac{1}{2\pi}\int_0^{2\pi}d\theta\ket{\psi_0}\bra{\psi_0}.
\end{equation}
The $K$th phase-randomized state would be
\begin{equation}
\label{eq:mixedsymmetricfromphaserandomization}
\begin{split}
\rho_K&=\frac{1}{2\pi}\int_0^{2\pi}d\theta\ket{\psi_K}\bra{\psi_K}=\frac{1}{2\pi}\int_0^{2\pi}d\theta\ U^K\ket{\psi_0}\bra{\psi_0}(U^K)^\dag\\
&=U^K\frac{1}{2\pi}\int_0^{2\pi}d\theta\ket{\psi_0}\bra{\psi_0}(U^K)^\dag=U^K\rho_0(U^K)^\dag,
\end{split}
\end{equation}
which shows that we end up with mixed symmetric states from phase-randomization. Substituting equations \eqref{eq:multimodesplitinphotonsubspaces} and \eqref{eq:unitaryoperatorpower} into \eqref{eq:mixedsymmetricfromphaserandomization}, we obtain
\begin{widetext}
\begin{equation}
\label{eq:kthphaserandomizedstate}
\begin{split}
\rho_K&=\frac{1}{2\pi}\int_0^{2\pi}d\theta\bigoplus_{N,N',R,R'=0}^\infty\sqrt{p_N p_{N'}}(U_R)^K\ket{\psi_{0,N}}\bra{\psi_{0,N'}}[(U_{R'})^K]^\dag\\
&=\frac{1}{2\pi}\int_0^{2\pi}d\theta\bigoplus_{N,N'=0}^\infty\sqrt{p_N p_{N'}}(U_N)^K\ket{\psi_{0,N}}\bra{\psi_{0,N'}}[(U_{N'})^K]^\dag\\
&=\bigoplus_{N,N'=0}^\infty(U_N)^K\frac{1}{2\pi}\int_0^{2\pi}d\theta\sqrt{p_N p_{N'}}\ket{\psi_{0,N}}\bra{\psi_{0,N'}}[(U_{N'})^K]^\dag
%\\&
=\bigoplus_{N,N'=0}^\infty(U_N)^KI_0^{(N,N')}[(U_{N'})^K]^\dag
\end{split}
\end{equation}
with
\begin{equation}
\label{eq:phaserandomizationintegral}
\begin{split}
I_0^{(N,N')}&=\frac{1}{2\pi}\int_0^{2\pi}d\theta\sqrt{p_N p_{N'}}\ket{\psi_{0,N}}\bra{\psi_{0,N'}}\\
&=\sum_{\substack{n_i=0,i\in[M]\\\sum_kn_k=N}}^N\sum_{\substack{n_j'=0,j\in[M]\\\sum_ln_l'=N'}}^{N'}A(\vec{r},\vec{n})A(\vec{r},\vec{n}')\times
%\\&\times
\frac{1}{2\pi}\int_0^{2\pi}d\theta B(\theta,\vec{n})(B(\theta,\vec{n}'))^*\ket{n_1,\ldots,n_M}\bra{n_1',\ldots,n_M'}.
\end{split}
\end{equation}
Because
%\Ioannis
{\begin{equation}
\label{eq:phaserandomizationdelta}
\begin{split}
\frac{1}{2\pi}\int_0^{2\pi}d\theta B(\theta,\vec{n})(B(\theta,\vec{n}'))^* %\\
&=\prod_{m=1}^Me^{i\Delta\theta_mn_m}e^{-i\Delta\theta_mn_m'}\frac{1}{2\pi}\int_0^{2\pi}d\theta e^{i\theta(\sum_pn_p-\sum_qn_q')}\\
&=\prod_{m=1}^Me^{i\Delta\theta_m(n_m-n_m')}\delta_{\sum_pn_p,\sum_qn_q'},
\end{split}
\end{equation}}
the $K$th density matrix becomes
%\Ioannis
{\begin{equation}
\label{eq:phaserandomizationNsplit}
\begin{split}
\rho_K&=\bigoplus_{N,N'=0}^\infty\sum_{\substack{n_i=0,\\ i\in[M]\\\sum_kn_k=N}}^N\sum_{\substack{n_j'=0,\\ j\in[M]\\\sum_ln_l'=N'}}^{N'}A(\vec{r},\vec{n})A(\vec{r},\vec{n}')\prod_{m=1}^M\delta_{\sum_pn_p,\sum_qn_q'}\times %\\&\times 
e^{i\Delta\theta_m(n_m-n_m')}(U_N)^K\ket{n_1,\ldots,n_M}\bra{n_1',\ldots,n_M'}[(U_{N'})^K]^\dag\\
&=\bigoplus_{N=0}^\infty\sum_{\substack{n_i=0,i\in[M]\\\sum_kn_k=N}}^N\sum_{\substack{n_j'=0,j\in[M]\\\sum_ln_l'=N}}^NA(\vec{r},\vec{n})A(\vec{r},\vec{n}')\times%\\&\times
\prod_{m=1}^Me^{i\Delta\theta_m(n_m-n_m')}(U_N)^K\ket{n_1,\ldots,n_M}\bra{n_1',\ldots,n_M'}[(U_N)^K]^\dag\\
&=\bigoplus_{N=0}^\infty p_N(U_N)^K\ket{\phi_{0,N}}\bra{\phi_{0,N}}[(U_N)^K]^\dag%\\&
=\bigoplus_{N=0}^\infty p_N\rho_{K,N}
\end{split}
\end{equation}}
with
\begin{equation}
\label{eq:phaserandomizationNpure}
\ket{\phi_{0,N}}=\frac{1}{\sqrt{p_N}}\sum_{\substack{n_i=0,i\in[M]\\\sum_kn_k=N}}^NA(\vec{r},\vec{n})\prod_{m=1}^Me^{i\Delta\theta_mn_m}\ket{n_1,\ldots,n_M},
\end{equation}
which shows that $\rho_K$ is a mixture of pure states with different total number of photons. %splits in total photon number subspaces.
\end{widetext}

Next, we show that the minimum-error measurement for the phase-randomized states is actually to perform the square-root measurement (the square-root measurement is not always optimal for mixed symmetric states). %One can easily show that 
The $L$ possible states are equiprobable, and are mixtures of pure states with different total photon numbers. The $L$ different pure states with a particular total photon number are clearly symmetric (because the symmetry operation does not change the total number of photons). Therefore the ``overall" square-root measurement is equivalent to first counting the total number of photons, and then performing the square-root measurement in each total photon number subspace. For equiprobable states, the measurement operators for the square-root measurement are given by 
\begin{equation}
\label{eq:generalmeasurementoperatorequiprobable}
\pi_i=\frac{1}{L}\rho^{-\frac{1}{2}}\rho_i\rho^{-\frac{1}{2}}.
\end{equation}
The average density operator can be written
\begin{align}
\label{eq:prioraveragedensitymatrix}
\rho&=\frac{1}{L}\sum_{i=0}^{L-1}\rho_i=\frac{1}{L}\sum_{i=0}^{L-1}\bigoplus_{N=0}^\infty p_N\rho_{i,N} \notag \\
&=\frac{1}{L}\bigoplus_{N=0}^\infty p_N\sum_{i=0}^{L-1}\rho_{i,N}=\bigoplus_{N=0}^\infty p_N\rho_N,
\end{align}
where $\rho_N$ is the average density matrix in the $N$-photon subspace.
Because the average density matrix $\rho$ is block diagonal, we can write
\begin{equation} 
\label{eq:minussquarerootdensitymatrix}
\rho^{-\frac{1}{2}}=\bigoplus_{N=0}^\infty p_N^{-\frac{1}{2}}(\rho_N)^{-\frac{1}{2}}.
\end{equation}
%\Erika{\bf The equation above, and similar places, should perhaps have a $\bigoplus$ as the summation sign instead of $\Sigma$?}
The measurement operator becomes
\begin{equation}
\label{eq:measurementoperatorsplit}
\begin{split}
\pi_i&=\frac{1}{L}\bigoplus_{N_1,N_2,N_3=0}^\infty p_{N_1}^{-\frac{1}{2}}p_{N_2}p_{N_3}^{-\frac{1}{2}}(\rho_{N_1})^{-\frac{1}{2}}\rho_{i,N_2}(\rho_{N_3})^{-\frac{1}{2}}\\
&=\frac{1}{L}\bigoplus_{N=0}^\infty(\rho_N)^{-\frac{1}{2}}\rho_{i,N}(\rho_N)^{-\frac{1}{2}}=\bigoplus_{N=0}^\infty\pi_i^{(N)}.
\end{split}
\end{equation}
We see that the square-root measurement operators can be written as a direct sum of operators acting in subspaces with different total photon numbers. The measurement operators in the $N$-photon subspace are given by
\begin{equation}
\label{eq:measurementoperatorNphotonsubspace}
\pi_i^{(N)}=\frac{1}{L}(\rho_N)^{-\frac{1}{2}}\rho_{i,N}(\rho_N)^{-\frac{1}{2}},
\end{equation}
with
%\Ioannis
{\begin{equation}
\label{eq:priordensitymatrixNsubspace}
\begin{split}
&\rho_{N}=\frac{1}{L}\sum_{l=0}^{L-1}\rho_{l,N}=\frac{1}{L}\sum_{l=0}^{L-1}\ket{\phi_{l,N}}\bra{\phi_{l,N}}, \\ &\ket{\phi_{l,N}}=(U_N)^l\ket{\phi_{0,N}}.
\end{split}
\end{equation}}
Eqn. \eqref{eq:measurementoperatorNphotonsubspace} is clearly the square-root measurement for the states $|\phi_{l,N}\rangle$ in the $N$-photon subspace. The overall square-root measurement for the phase-randomized states is therefore the same as the combination of the square-root measurements for distinguishing between the resulting pure symmetric states in each photon-number subspace. The square-root measurement is known to be optimal for distinguishing between pure symmetric equiprobable states. As we already mentioned, the optimal measurement for the phase-randomized states can be realized by first determining the total photon number, and then distinguishing between the resulting pure symmetric states in the corresponding photon-number subspace. We have verified that this is the same as the overall square-root measurement for the mixed phase-randomized states in \cref{eq:generalmeasurementoperatorequiprobable}.

\section{The optimal measurement for the two-mode states in eq. \eqref{eq:twomixed}}
\label{app:twomode}

To obtain the optimal measurement for distinguishing between the two-mode phase-randomized coherent states in eq. \eqref{eq:twomixed}, we note that applying the same unitary transform to all states we want to distinguish between leaves the optimal success probability unchanged. Since a unitary operation corresponds to a basis change, the optimal measurement operators for the original states can be found from the measurement operators for the transformed states, using the inverse of the unitary transform. 
We will use the unitary transform
$U_{\rm BS} \ket{\alpha, \beta} = \ket{(\alpha+\beta)/\sqrt 2, (\alpha-\beta)/\sqrt 2}$, 
%where $U_\text{BS}$ is 
corresponding to a $50$/$50$ beam splitter,
giving
\begin{equation} \label{eq:50/50beamsplittertwomodestates}
\begin{split}
%U_\text{BS}\ket{\psi_0}\bra{\psi_0}U_\text{BS}^\dag & =U_\text{BS}\ket{\alpha,\alpha}\bra{\alpha,\alpha}U_\text{BS}^\dag=|\sqrt{2}\alpha,0\rangle\langle\sqrt{2}\alpha,0| \\ 
U_\text{BS}\ket{\psi_0}& =U_\text{BS}\ket{\alpha,\alpha}=|\sqrt{2}\alpha,0\rangle\\ 
U_\text{BS}\ket{\psi_1}& =U_\text{BS}\ket{\alpha,-\alpha}=|0,\sqrt{2}\alpha\rangle .
%U_\text{BS}\ket{\psi_1}\bra{\psi_1}U_\text{BS}^\dag & =U_\text{BS}\ket{\alpha,-\alpha}\bra{\alpha,-\alpha}U_\text{BS}^\dag=|0,\sqrt{2}\alpha\rangle\langle0,\sqrt{2}\alpha|
\end{split}
\end{equation}
This means that we need to distinguish between the states 
%\Ioannis
{\begin{equation} \label{eq:beamsplitterphaserandomizedtwomodestates}
\begin{split}
\rho_0'=U_\text{BS}\rho_0U_\text{BS}^\dag & = \frac{1}{2\pi}\int_0^{2\pi}d\theta\;U_\text{BS}\ket{\psi_0}\bra{\psi_0}U_\text{BS}^\dag \\ & = e^{-2\abs{\alpha}^2} \sum_{N=0}^\infty{\frac{2^N\abs{\alpha}^{2N}}{N!}\ket{N,0}\bra{N,0}}, \\
\rho_1'=U_\text{BS}\rho_1U_\text{BS}^\dag & = \frac{1}{2\pi}\int_0^{2\pi}d\theta\;U_\text{BS}\ket{\psi_1}\bra{\psi_1}U_\text{BS}^\dag \\ & = e^{-2\abs{\alpha}^2} \sum_{N=0}^\infty{\frac{2^N\abs{\alpha}^{2N}}{N!}\ket{0,N}\bra{0,N}}.
\end{split}
\end{equation}}
We will also find that the square-root measurement is equal to the Helstrom measurement in this particular case, if the two states are equiprobable, with only minor differences otherwise.
Using the density matrices in \cref{eq:beamsplitterphaserandomizedtwomodestates}, we obtain
%\Ioannis
\begin{align} \label{eq:beamsplitterdifferenceoperator}
A'&=p_0\rho_0'-p_1\rho_1' \notag \\ &=e^{-2\abs{\alpha}^2} \sum_{N=0}^\infty\frac{2^N\abs{\alpha}^{2N}}{N!}\big(p_0\ket{N,0}\bra{N,0}\\ 
&~~~~~~~~~~~~~~~~~~~~~~~~~~~~~
-p_1\ket{0,N}\bra{0,N}\big).\notag
\end{align}
Because the states $\ket{N,0}$ and $\ket{0,N}$ are orthogonal for $N\neq0$ and only overlap for $N=0$, the optimal measurement operators for the Helstrom measurement are
\begin{equation} \label{eq:optimalmeasurementphaserandomizedtwomodecase1}
\begin{split}
\Pi_0 & =\sum_{N=1}^\infty{\ket{N,0}\bra{N,0}}+\ket{0,0}\bra{0,0} \\
\Pi_1 & =\sum_{N=1}^\infty{\ket{0,N}\bra{0,N}}%+p_1\ket{0,0}\bra{0,0}
\end{split}
\end{equation}
if $p_0>p_1$, and 
\begin{equation} \label{eq:optimalmeasurementphaserandomizedtwomodecase2}
\begin{split}
\Pi_0 & =\sum_{N=1}^\infty{\ket{N,0}\bra{N,0}} \\
\Pi_1 & =\sum_{N=1}^\infty{\ket{0,N}\bra{0,N}}+\ket{0,0}\bra{0,0}
\end{split}
\end{equation}
if $p_0<p_1$. If $p_0=p_1$, then the projector onto $\ket{0,0}$ can be ``split up" in any proportion between $\Pi_0$ and $\Pi_1$.
The above measurement operators add up to identity on the support of %in terms of the basis $\{\ket{0,0}, \ket{N,0}, \ket{0,N}\}$ of 
$A'$. (States $\ket{j,k}$ where both $j\neq 0$ and $k\neq 0$ are not included in the support of $A'$, $\rho_0'$ or $\rho_1'$; corresponding events will never occur in an ideal setup. Should experimental imperfections be present, it is also possible to derive optimal measurements for this situation. The measurement one should aim to realize may or may not be identical to the optimal measurement for ideal conditions~\cite{Erika_imperfect}.)
%Other possible measurement events can be given by
%\begin{equation} \label{eq49}
%\Pi_0=\mathds{1}-\Pi_+-\Pi_-=\sum_{j,k>0}^\infty\ket{j,k}\bra{j,k}
%\end{equation}
%where $\Pi_0$ corresponds to the measurement operator in the subspace $\lambda_i=0$ of the operator $A'$. 
%\Erika{\bf The states are labelled with 0 and 1, but the measurement operators with $+$ and $-$. Except the measurement operator that corresponds to results that never occur... This is a little confusing. It might be better to let the measurement operator corresponding to $\rho_0$ be $\Pi_0$, and $\Pi_1$ for $\rho_1$.}
%Because $\Pi_0\rho_i'=0$ for $i=0,1$, the operator $\Pi_0$ is not included (has zero contribution) in the measurement process of the two density matrices.

To show that the square-root measurement coincides with the above Helstrom measurement for the above two-mode phase-randomized coherent states if they are equiprobable, we derive the measurement operators for the square-root measurement. %\Erika{\bf If we wish, it should be possible to show this in a general case with simpler notation (shorter equations).  The two measurements (Helstrom and square-root) should be the same whenever the two possible states have support in completely distinct parts of the Hilbert space.} 
We can write $\rho_0'$ and $\rho_1'$ as
\begin{equation} \rho_0'=\sum_{i=0}^\infty c_iP_i,\quad \rho_1'=\sum_{i=0}^\infty c_iQ_i
\end{equation}
where %$P_i,Q_i$ are rank-$1$ projectors \Giannis{as the probabilities $c_i$ are non-degenerate} \Erika{Can we assume non-degeneracy and do we need it? Can we simply not state that $P_i, Q_i$ should be rank 1, since such a decomposition is always possible even when some of the probabilities are equal?} \Giannis{Yes. I just said that because that holds for our case. Even if the case was degenerate, the same proof holds with the projectors having the rank of the eigenvalue degeneracy.},
$P_0=Q_0$, and $P_iQ_j=0$ for $i,j>0$. We write
\begin{equation}
\rho'_0=c_0P_0+\tilde{\rho}_0,\quad\rho_1'=c_0P_0+\tilde{\rho}_1
\end{equation}
and
\begin{equation}
\rho'=p_0\rho_0'+p_1\rho_1'=c_0P_0+p_0\tilde{\rho}_0+p_1\tilde{\rho}_1
\end{equation}
because $p_0+p_1=1$. We have $\tilde{\rho}_0,\tilde{\rho}_1\succ0$ because $c_i>0$. Then %\Giannis{I changed the symbol from $>$ to $\succ$} Then
\begin{equation}
(\rho')^{-\frac{1}{2}}=c_0^{-\frac{1}{2}}P_0+p_0^{-\frac{1}{2}}\tilde{\rho}_0^{-\frac{1}{2}}+p_1^{-\frac{1}{2}}\tilde{\rho}_1^{-\frac{1}{2}}.
\end{equation}
The first measurement operator is
%\Ioannis
{\begin{align}
\pi_0'&=p_0(\rho')^{-\frac{1}{2}}\rho_0'(\rho')^{-\frac{1}{2}} \notag \\
&=p_0\Big[c_0^{-\frac{1}{2}}P_0+p_0^{-\frac{1}{2}}\tilde{\rho}_0^{-\frac{1}{2}}+p_1^{-\frac{1}{2}}\tilde{\rho}_1^{-\frac{1}{2}}\Big](c_0P_0+\tilde{\rho}_0)\times \notag \\ &\times\Big[c_0^{-\frac{1}{2}}P_0+p_0^{-\frac{1}{2}}\tilde{\rho}_0^{-\frac{1}{2}}+p_1^{-\frac{1}{2}}\tilde{\rho}_1^{-\frac{1}{2}}\Big] \notag \\
&=p_0P_0+\tilde{\rho}_0^{-\frac{1}{2}}\tilde{\rho}_0\tilde{\rho}_0^{-\frac{1}{2}} %=p_0P_0+\sum_{i,j,k=1}^\infty c_i^{-\frac{1}{2}}c_jc_k^{-\frac{1}{2}}P_iP_jP_k \notag \\&
=p_0P_0+\sum_{i=1}^\infty P_i \notag \\ &=p_0\ket{0,0}\bra{0,0}+\sum_{i=1}^\infty\ket{i,0}\bra{i,0}.
\end{align}}
Using similar arguments, we obtain 
\begin{equation}
\label{eq:secondsquarerootoperatorphaserandomizedtwomode}
\pi_1'=p_1P_0+\sum_{i=1}^\infty Q_i=p_1\ket{0,0}\bra{0,0}+\sum_{i=1}^\infty\ket{0,i}\bra{0,i}.
\end{equation}
That is, if $p_0=p_1$, then the square-root measurement is a minimum-error measurement for the mixed states in \eqref{eq:twomixed}. When $p_0\neq p_1$, the only difference between the square-root and the minimum-error measurement is in how the projector $\ket 0 \bra 0$ is ``shared" between the two measurement operators.
\section{Calculating pairwise overlaps between $\ket{\psi_{bc,N}}$ and Gram matrix elements}
\label{app:threemode}

From the trinomial expansion
\begin{align}
\label{eq:trinomialexpansion}
(a+b+c)^N&=\sum_{\substack{j,k,l=0\\ j+k+l=N}}^N\binom{N}{j,k,l}a^jb^kc^l \notag \\
&=\sum_{\substack{j,k,l=0\\ j+k+l=N}}^N\frac{N!}{j!k!l!}a^jb^kc^l
\end{align}
%\begin{equation} 
%\label{eq:trinomialexpansion}
%\begin{align}
%(a+b+c)^N&=\sum_{\substack{j,k,l=0\\ j+k+l=N}}^N\binom{N}{j,k,l}a^jb^kc^l\\
%&=\sum_{\substack{j,k,l=0\\ j+k+l=N}}^N\frac{N!}{j!k!l!}a^jb^kc^l
%\end{align}
%\end{equation}
for $a=b=c=1$, we find the normalization factor $M_N$ of the states $\ket{\psi_{bc,N}}$ in \cref{eq:threemodeNpure}
\begin{equation} \label{eq:trinomialexpansionthreemode}
3^N=\sum_{\substack{j,k,l=0\\ j+k+l=N}}^N\frac{N!}{j!k!l!}\Rightarrow M_N=\sum_{\substack{j,k,l=0\\ j+k+l=N}}^N\frac{1}{j!k!l!}=\frac{3^N}{N!}.
\end{equation}
For the elements of the Gram matrices in \cref{eq:grammatrixNthreemode,eq:grammatrixphaseencodedNpositive}, we used the multinomial expansion
\begin{equation} \label{eq:multinomialexpansion}
(x_1+x_2+\cdots+x_k)^N=\sum_{\substack{n_1,n_2,\ldots\\ \ldots,n_k=0\\ n_1+n_2+\ldots\\ \ldots +n_k=N}}^N\frac{N!}{n_1!n_2!\cdots n_k!}x_1^{n_1}x_2^{n_2}\cdots x_k^{n_k}.
\end{equation}
For example, for the four-mode coherent states in the subspace with $N>0$ photons, we have
\begin{equation}
\label{eq:grammatrixelementfourmode}
\begin{split}(\tilde{\mathcal{G}}_N)_{0,1}=\langle\tilde{\psi}_{00,N}|\tilde{\psi}_{01,N}\rangle &=\frac{1}{\tilde{M}_N}\sum_{\substack{j,k,l,m=0\\ j+k+l+m=N}}^N\frac{(-1)^l(-1)^m}{j!k!l!m!} \\
&=\frac{1}{\frac{4^N}{N!}}\frac{(1+1-1-1)^N}{N!}=0.
\end{split}
\end{equation}

\section{Success probability and cheating probability for quantum oblivious transfer}
\label{app:protsuccheatprob}
%\Erika{\bf Need a shorter section title}
When calculating the probabilities in \cref{eq:protocolsucprobthreemode,eq:protocolsucprobfourmode,eq:protocolsucprobphaseencoded}, for distinguishing between pairs of states, selected from four symmetric pure states, it is convenient to express the states using the ``quantum Fourier transform" orthonormal basis $\ket{A_i}$~\cite{reichmuth2024incompletequantumoblivioustransfer}. 
% \Giannis{one sided paper not in arxiv. I changed the title and date in Refs to the latest version that Ittoop sent me}. 
The Gram matrix for a set of pure states is circulant, and circulant matrices are diagonal when expressed in this basis. This method
%The method of calculating the success probability of the protocol using the ``Quantum Fourier Transform" orthonormal basis $\ket{A_i}$~\cite{OneSided2021} and not representing the states in the multi-mode photon number state basis
can be used also for more than four states, e.g. when we encode three bits of classical information in eight symmetric quantum states. The quantum Fourier transform is a unitary matrix $\mathcal{F}$ acting on a set of basis state vectors. The matrix $\mathcal{F}$ is the discrete Fourier transform, given by
\begin{equation}
\label{eq:quantumfourier}
\mathcal{F}=\frac{1}{\sqrt{L}}\begin{pmatrix}
1 & 1 & 1 & 1 & \cdots & 1 \\
1 & \omega & \omega^2 & \omega^3 & \cdots & \omega^{L-1} \\
1 & \omega^2 & \omega^4 & \omega^6 & \cdots & \omega^{2(L-1)} \\
1 & \omega^3 & \omega^6 & \omega^9 & \cdots & \omega^{3(L-1)} \\
\vdots & \vdots & \vdots & \vdots & \ddots & \vdots \\
1 & \omega^{L-1} & \omega^{2(L-1)} & \omega^{3(L-1)} & \cdots & \omega^{(L-1)^2}
\end{pmatrix}
\end{equation}
where $\omega=e^{\frac{2\pi i}{L}}$ with $L=2^n$. For us, $n$ is the number of classical bits that is encoded in the set of quantum states. A circulant matrix of size $L\times L$ is generated from an $L$-vector $\{c_0,c_1,\ldots,c_{L-1}\}$ by cyclically permuting its entries. The matrix is given by
\begin{equation}
\textbf{C}=\begin{pmatrix}
c_0 & c_1 & \cdots & c_{L-1} \\
c_{L-1} & c_0 & \cdots & c_{L-2} \\
\vdots & \vdots & \ddots & \vdots \\
c_1 & c_2 & \cdots & c_0
\end{pmatrix}.
\end{equation}
Here, each row is a right cyclic shift of the row above it. The circulant operator $\text{circ}(\cdot)$ creates a circulant matrix from the vector $\{c_0,c_1,\ldots,c_{L-1}\}$, so that
\begin{equation}
\textbf{C}=\text{circ}(c_0,c_1,\ldots,c_{L-1}).
\end{equation}
The element at position $(i,k)$ of the matrix is given by $(\textbf{C})_{ik}=c_{(k-i)\text{mod}L}$ where $i,k$ range from $0$ to $L-1$. The Gram matrix of the states $\ket{\psi_i}=U^i\ket{\psi_0}$ is circulant with elements $c_l=\mathcal{G}_{0l}$ and has the form
\begin{align}
\label{eq:Grammatrixcirculant}
\mathcal{G}&=\text{circ}\big(\mathcal{G}_{00}, \mathcal{G}_{01}, \mathcal{G}_{02},\ldots, \mathcal{G}_{0L-1}\big) \notag \\
&=\begin{pmatrix}
\mathcal{G}_{00} & \mathcal{G}_{01} & \cdots & \mathcal{G}_{0L-2} & \mathcal{G}_{0L-1} \\
\mathcal{G}_{0L-1} & \mathcal{G}_{00} & \cdots & \mathcal{G}_{0L-3} & \mathcal{G}_{0L-2} \\
\vdots & \vdots & \ddots & \vdots & \vdots \\
\mathcal{G}_{02} & \mathcal{G}_{03} & \cdots & \mathcal{G}_{00} & \mathcal{G}_{01} \\
\mathcal{G}_{01} & \mathcal{G}_{02} & \cdots & \mathcal{G}_{0L-1} & \mathcal{G}_{00}
\end{pmatrix}.
\end{align}
We follow the notation $\ket{\psi_i}=U^i\ket{\psi_0}$ where $\ket{\psi_0}\equiv\ket{\psi_{00\ldots}}$ for the set of symmetric states. It can be proven that the eigenvalues of a circulant matrix are~\cite{CirculantMatr}
\begin{equation}
\lambda_k=\sum_{l=0}^{L-1}c_l\omega^{kl}
\end{equation}
with $\omega=e^{\frac{2\pi i}{L}}$ and $k=0,1,\ldots,L-1$.
\begin{widetext}
Therefore, the eigenvalues of the Gram matrix can be written as
\begin{align}
\label{eq:eigenvalueequationgrammatrix}
\lambda_k&=\sum_{l=0}^{L-1}\mathcal{G}_{0l}\omega^{kl}=\sum_{l=0}^{L-1}\mathcal{G}_{0l}\frac{1}{L}L\omega^{kl} \notag \\
&=\mathcal{G}_{00}\frac{1}{L}L\omega^{k\cdot0}+\mathcal{G}_{01}\frac{1}{L}L\omega^{k\cdot1}+\cdots+\mathcal{G}_{0L-1}\frac{1}{L}L\omega^{k\cdot(L-1)} \notag \\
&=\bigg[\mathcal{G}_{00}\frac{1}{L}\omega^{k(0-0)}+\mathcal{G}_{11}\frac{1}{L}\omega^{k(1-1)}+\cdots+\mathcal{G}_{L-1,L-1}\frac{1}{L}\omega^{k(L-1-L+1)}\bigg] \notag \\
&+\bigg[\mathcal{G}_{01}\frac{1}{L}\omega^{k(1-0)}+\mathcal{G}_{12}\frac{1}{L}\omega^{k(2-1)}+\cdots+\mathcal{G}_{L-2,L-1}\frac{1}{L}\omega^{k(L-1-L+2)}+\mathcal{G}_{L-1,0}\frac{1}{L}\omega^{k(0-L+1)}\bigg] \notag \\
&+\cdots \notag \\
&+\bigg[\mathcal{G}_{0L-1}\frac{1}{L}\omega^{k(L-1-0)}+\mathcal{G}_{10}\frac{1}{L}\omega^{k(0-1)}+\cdots+\mathcal{G}_{L-1,L-2}\frac{1}{L}\omega^{k(L-2-L+1)}\bigg] \notag \\
&=\sum_{j,m=0}^{L-1}\mathcal{G}_{mj}\frac{1}{L}\omega^{k(j-m)}=\sum_{j,m=0}^{L-1}\mathcal{F}_{mk}^*\mathcal{G}_{mj}\mathcal{F}_{jk} \notag \\
&=\sum_{j,m=0}^{L-1}(\mathcal{F}^\dag)_{km}\mathcal{G}_{mj}\mathcal{F}_{jk} \notag
=(\mathcal{F}^\dag\mathcal{G}\mathcal{F})_{kk}=(\mathcal{F}^{-1}\mathcal{G}\mathcal{F})_{kk}
\end{align}
where we used the properties
\begin{equation}
\mathcal{G}_{ij}=\mathcal{G}_{(i+m)\text{mod}L,(j+m)\text{mod}L},\quad\mathcal{F}^{-1}=\mathcal{F}^\dag
\end{equation}
with $k=0,1,\ldots,L-1$ and also $\omega^{kL}=1$. 
Then, we see that
\begin{equation}
\label{eq:eigenvaluesoverlapstates}
\lambda_k=\sum_{j,m=0}^{L-1}\mathcal{F}_{km}^*\mathcal{F}_{kj}\mathcal{G}_{mj}=\sum_{j,m=0}^{L-1}\bra{\psi_m}\mathcal{F}_{km}^*\mathcal{F}_{kj}\ket{\psi_j}=\bra{B_k}\ket{B_k}
\end{equation}
with
\begin{equation}
\label{eq:unormalizedeigenvaluesstates}
\ket{B_k}=\sum_{j=0}^{L-1}\mathcal{F}_{kj}\ket{\psi_j}.
\end{equation}
The states $\ket{B_k}$ are unnormalized. We will show that these states are orthogonal.
\end{widetext}

We have
\begin{align}
\label{eq:orthogonalityofunormalizedstatesone}
\bra{B_l}\ket{B_k}&=  \sum_{j,m=0}^{L-1}\mathcal{F}_{lm}^*\mathcal{F}_{kj}\bra{\psi_m}\ket{\psi_j}=\sum_{j,m=0}^{L-1}\mathcal{F}_{lm}^*\mathcal{F}_{kj}\mathcal{G}_{mj} \notag \\
&=\sum_{j,m=0}^{L-1}\frac{1}{L}\omega^{kj-lm}\mathcal{G}_{mj}=\sum_{i=0}^{L-1}\frac{1}{L}\mathcal{G}_{0i}\omega^{ki}\sum_{m=0}^{L-1}\omega^{(k-l)m},
\end{align}
and for $k\neq l$ it holds that
\begin{equation}
\label{eq:orthogonalityofunormalizedstatestwo}
\bra{B_l}\ket{B_k}=\sum_{i=0}^{L-1}\frac{1}{L}\mathcal{G}_{0i}\omega^{ki}\frac{1-\omega^{(k-l)L}}{1-\omega^{k-l}}=0.
\end{equation}
Therefore
\begin{equation}
\label{eq:orthogonalityofunormalizedstatesthree}
\bra{B_l}\ket{B_k}=\lambda_k\delta_{lk}.
\end{equation}
We define an orthonormal basis $\ket{A_k}$ from $\ket{B_k}$ as
\begin{equation}
\label{eq:orthonormalbasis}
\ket{B_k}=\sqrt{\lambda_k}\ket{A_k}=\sum_{j=0}^{L-1}\mathcal{F}_{kj}\ket{\psi_j}.
\end{equation}
%\Ioannis{The orthonormal basis states $\ket{A_i}$ are given by
%\begin{equation}
%\label{eq:quantumfourierbasisstates}
%\sqrt{\lambda_i}\ket{A_i}=\sum_{j=0}^{L-1}\mathcal{F}_{ij}\ket{\psi_j}.
%\end{equation}}
We express the states $\ket{\psi_j}$ in the basis $\{\ket{A_k}\}$ as 
%We start from \Erika{\bf We could shorten this if we wanted -- it is clear that we should use the inverse Fourier transform.}
%\begin{align}
%\label{eq:statesinorthonormalbasis}
%\sqrt{\lambda_i}\ket{A_i}&=\sum_{j=0}^{L-1}\mathcal{F}_{ij}\ket{\psi_j}\Rightarrow \notag \\
%\sum_{i=0}^{L-1}\mathcal{F}_{im}^*\sqrt{\lambda_i}\ket{A_i}&=\sum_{j=0}^{L-1}\sum_{i=0}^{L-1}\mathcal{F}_{ij}\mathcal{F}_{im}^*\ket{\psi_j} \notag \\
%=\sum_{j=0}^{L-1}\sum_{i=0}^{L-1}(\mathcal{F}^\dag)_{mi}\mathcal{F}_{ij}\ket{\psi_j}&=\sum_{j=0}^{L-1}\delta_{mj}\ket{\psi_j}\Rightarrow \notag \\
%\ket{\psi_m}=\sum_{i=0}^{L-1}&\mathcal{F}_{im}^*\sqrt{\lambda_i}\ket{A_i}
%\end{align}
\begin{equation}
\label{eq:statesinorthonormalbasis}
\ket{\psi_m}=\sum_{i=0}^{L-1}\mathcal{F}_{im}^*\sqrt{\lambda_i}\ket{A_i}    
\end{equation}
where we used the unitarity relation of $\mathcal{F}$.
%\begin{equation}
%\label{eq:unitarityofquantumfouriertransform}
%\mathcal{F}^\dag\mathcal{F}=\mathds{1}\Rightarrow\sum_{i=0}^{L-1}(\mathcal{F}^\dag)_{mi}\mathcal{F}_{ij}=\delta_{mj}
%\end{equation}
We denote the corresponding density matrices by
\begin{equation}
\label{eq:firstdensitymatrixnotation}
\sigma_j=\ket{\psi_j}\bra{\psi_j},
\end{equation}
or, writing out the indices explicitly,
\begin{equation}
\label{eq:seconddensitymatrixnotation}
\sigma_{x_0x_1\ldots x_{n-1}}=|\psi_{x_0x_1\ldots x_{n-1}}\rangle\langle\psi_{x_0x_1\ldots x_{n-1}}|.
\end{equation}
If Bob is honest and wishes to learn the value of the $k^{\rm th}$ bit, then he performs a measurement to distinguish between the sets of states $\{\sigma_{x_0x_1\ldots x_{n-1}}|_{x_k=0}\}$ and $\{\sigma_{x_0x_1\ldots x_{n-1}}|_{x_k=1}\}$ with $x_i\in\{0,1\}$ for $i\neq k$,
where all states are equiprobable. This is the same
as making a measurement to distinguish between the
equiprobable states
\begin{equation}
\label{eq:nbitstates}
\rho_{b}=\frac{1}{2^{n-1}}\sum_{\substack{x_i\in\{0,1\}\\i\neq k}}\sigma_{x_0x_1\ldots x_{n-1}}|_{x_k=b}
\end{equation}
with $b=0,1$. If we construct the difference operator $A=\frac{1}{2}(\rho_0-\rho_1)$ for the equiprobable states in \cref{eq:nbitstates}, then we find the probability for correctly identifying the value of the $k$th classical bit using equation \eqref{eq:helstrom} as
%\begin{equation}
%\label{eq:probkclassicalbit}
%P_{x_k}=\frac{1}{2}\Big[1+\Tr|A|\Big].
%\end{equation}
\begin{equation}
\label{eq:probkclassicalbit}
P_{x_k}=\frac{1}{2}+\Tr[A\Pi_0]=\frac{1}{2}+\sum_{\lambda_i\geq0}\lambda_i.
\end{equation}
where $\lambda_i$ are the eigenvalues of $A$. On average, Bob's probability to obtain a correct bit
value, which corresponds to the protocol success probability, is
\begin{equation}
\label{eq:protocolsuccessprobabilitymeanvalueofbits}
P_{\text{1bit}}=\frac{1}{n}\sum_{k=0}^{n-1}P_{x_k}.
\end{equation}
If Bob's success probability is independent of which bit he tries to obtain, then $P_{\text{1bit}}=P_{x_k}$ for every $k$.
%\Erika{\bf Haven't we already included the derivation below, of the Helstrom measurement, in the paper somewhere else (in the main paper)? Once is enough.} \Ioannis{ \bf We have to prove equation $(D23)$ because we use it when we examine the $L=4$ example. That's why i included it.}
%To start with, we prove \cref{eq:probkclassicalbit}. From \cref{eq:helstrom}, the probability for correctly identifying the state as $\rho_0$ or $\rho_1$ can be expressed as
%\begin{equation}
%\label{eq:probabilitycorrectalternativeexpression}
%P_{\text{corr}}=p_1+\Tr[A\Pi_0]=p_0-\Tr[A\Pi_1]
%\end{equation}
%where $A=p_0\rho_0-p_1\rho_1$. The optimal measurement operators are then
%\begin{equation}
%\label{eq:optimaloperatorseigenvectorsofA}
%\Pi_0=\sum_{\lambda_i\geq0}\ket{\lambda_i}\bra{\lambda_i},\quad\Pi_1=\sum_{\lambda_i<0}\ket{\lambda_i}\bra{\lambda_i}
%\end{equation}
%where $\ket{\lambda_i}$ is the eigenvector of $A$ with eigenvalue $\lambda_i$. We obtain
%\begin{equation}
%\label{eq:probabilitycorrectabsolutevaluematrixA}
%\begin{split}
%&P_{\text{corr}}=p_1+\sum_{\lambda_i\geq0}\lambda_i=p_0-\sum_{\lambda_i<0}\lambda_i\Rightarrow\\
%&2P_{\text{corr}}=p_0+p_1+\sum_{\lambda_i\geq0}\lambda_i-\sum_{\lambda_i<0}\lambda_i=1+\sum_{\lambda_i}\abs{\lambda_i}\Rightarrow\\
%&P_{\text{corr}}=\frac{1}{2}\Big[1+\Tr|A|\Big].
%\end{split}
%\end{equation}
Next, we apply the above result to the case of two classical bits using \cref{eq:nbitstates}.
We reproduce the analysis in~\cite{reichmuth2024incompletequantumoblivioustransfer} when four states encode two classical bits. If Bob is honest and wishes to learn the value of the first bit, then he performs a measurement to distinguish between the sets of states $\{\sigma_{00},\sigma_{01}\}$ and $\{\sigma_{10},\sigma_{11}\}$,
where all states are equiprobable. This is the same
as making a measurement to distinguish between the
equiprobable states $\rho_0=\frac{1}{2}(\sigma_{00}+\sigma_{01})$ and $\rho_1=\frac{1}{2}(\sigma_{10}+\sigma_{11})$. The probability of a correct result in this case is therefore
\begin{equation}
\label{eq:protocolsuccessprobabilitytwoclassicalbits}
P_{\text{corr}}\equiv P_{\text{1bit}}=\frac{1}{2}+\frac{1}{8}\Tr|\sigma_{00}+\sigma_{01}-\sigma_{11}-\sigma_{10}|.
\end{equation}
The $L=4$ discrete Fourier transform matrix is
\begin{equation}
\label{eq:quantumfouriertwoclassicalbits}
\mathcal{F}=\frac{1}{2}\begin{pmatrix}
1 & 1 & 1 & 1 \\
1 & i & -1 & -i \\
1 & -1 & 1 & -1 \\
1 & -i & -1 & i
\end{pmatrix}.
\end{equation}
\begin{widetext}
From \cref{eq:statesinorthonormalbasis} we have
\begin{equation}
\label{eq:statesinquantumfourierbasistwoclassicalbits}
\begin{split}
\ket{\psi_0}&\equiv\ket{\psi_{00}}=\frac{1}{2}(\sqrt{\lambda_0}\ket{A_0}+\sqrt{\lambda_1}\ket{A_1}+\sqrt{\lambda_2}\ket{A_2}+\sqrt{\lambda_3}\ket{A_3})\\
\ket{\psi_1}&\equiv\ket{\psi_{01}}=\frac{1}{2}(\sqrt{\lambda_0}\ket{A_0}-i\sqrt{\lambda_1}\ket{A_1}-\sqrt{\lambda_2}\ket{A_2}+i\sqrt{\lambda_3}\ket{A_3})\\
\ket{\psi_2}&\equiv\ket{\psi_{11}}=\frac{1}{2}(\sqrt{\lambda_0}\ket{A_0}-\sqrt{\lambda_1}\ket{A_1}+\sqrt{\lambda_2}\ket{A_2}-\sqrt{\lambda_3}\ket{A_3})\\
\ket{\psi_3}&\equiv\ket{\psi_{10}}=\frac{1}{2}(\sqrt{\lambda_0}\ket{A_0}+i\sqrt{\lambda_1}\ket{A_1}-\sqrt{\lambda_2}\ket{A_2}-i\sqrt{\lambda_3}\ket{A_3}).
\end{split}
\end{equation}
In the basis $\{\ket{A_0},\ket{A_1},\ket{A_2},\ket{A_3}\}$, we have
\begin{equation}
\label{eq:densitymatricestwoclassicalbitsquantumfourierbasis}
\begin{split}
\rho_0&=\frac{1}{2}(\sigma_{00}+\sigma_{01})=\frac{1}{8}\begin{pmatrix}
2\lambda_0 & \sqrt{\lambda_{01}}(1+i) & 0 & \sqrt{\lambda_{03}}(1-i) \\
\sqrt{\lambda_{01}}(1-i) & 2\lambda_1 & \sqrt{\lambda_{12}}(1+i) & 0 \\
0 & \sqrt{\lambda_{12}}(1-i) & 2\lambda_2 & \sqrt{\lambda_{23}}(1+i) \\
\sqrt{\lambda_{03}}(1+i) & 0 & \sqrt{\lambda_{23}}(1-i) & 2\lambda_3 
\end{pmatrix},\\
\rho_1&=\frac{1}{2}\big(\sigma_{11}+\sigma_{10}\big)=\frac{1}{8}\begin{pmatrix}
2\lambda_0 & -\sqrt{\lambda_{01}}(1+i) & 0 & -\sqrt{\lambda_{03}}(1-i) \\
-\sqrt{\lambda_{01}}(1-i) & 2\lambda_1 & -\sqrt{\lambda_{12}}(1+i) & 0 \\
0 & -\sqrt{\lambda_{12}}(1-i) & 2\lambda_2 & -\sqrt{\lambda_{23}}(1+i) \\
-\sqrt{\lambda_{03}}(1+i) & 0 & -\sqrt{\lambda_{23}}(1-i) & 2\lambda_3 
\end{pmatrix},\\
A&=\frac{1}{4}\big(\sigma_{00}+\sigma_{01}-\sigma_{11}-\sigma_{10}\big)\\
&=\frac{1}{8}\begin{pmatrix}
0 & \sqrt{\lambda_{01}}(1+i) & 0 & \sqrt{\lambda_{03}}(1-i) \\
\sqrt{\lambda_{01}}(1-i) & 0 & \sqrt{\lambda_{12}}(1+i) & 0 \\
0 & \sqrt{\lambda_{12}}(1-i) & 0 & \sqrt{\lambda_{23}}(1+i) \\
\sqrt{\lambda_{03}}(1+i) & 0 & \sqrt{\lambda_{23}}(1-i) & 0 
\end{pmatrix},
\end{split}
\end{equation}
where we have used the shorthand $\lambda_{ij}= \lambda_i\lambda_j$. The eigenvalue equation for the last matrix above is given by
$|A-\Lambda\mathds{1}|=0$, which, when evaluating the determinant, gives
\begin{equation}
\label{eq:determinantdifferencematrixtwoclassicalbits}
(8\Lambda)^4-2(8\Lambda)^2(\lambda_0+\lambda_2)(\lambda_1+\lambda_3)+16\lambda_{0123}=0,
\end{equation}
where $\lambda_{0123}=\lambda_0\lambda_1\lambda_2\lambda_3$. The four eigenvalues of the last matrix above are consequently given by the solutions of \cref{eq:determinantdifferencematrixtwoclassicalbits}, 
\begin{equation}
\label{eq:solutionstwoclassicalbits}
\Lambda_\pm^2=\frac{1}{8^2}\bigg[(\lambda_0+\lambda_2)(\lambda_1+\lambda_3)\ \pm\ \sqrt{(\lambda_0+\lambda_2)^2(\lambda_1+\lambda_3)^2-16\lambda_0\lambda_1\lambda_2\lambda_3}\ \bigg].
\end{equation}
From \cref{eq:eigenvaluesoverlapstates}, we find 
\begin{equation}
\label{eq:eigenavluesofgeneralgrammatrix}
\begin{split}
\lambda_0&=1+F+G+F^*=1+G+2\text{Re}F\\
\lambda_1&=1+iF-G-iF^*=1-G-2\text{Im}F\\
\lambda_2&=1-F+G-F^*=1+G-2\text{Re}F\\
\lambda_3&=1-iF-G+iF^*=1-G+2\text{Im}F
\end{split}
\end{equation}
in that order. The solutions can be written
\begin{equation}
\label{eq:solutionsintermsofpairwiseoverlaps}
\Lambda_\pm^2=\frac{1}{8^2}\Big[4(1-G^2)\ \pm\ 8\sqrt{(1+G)^2(\text{Im}F)^2+(1-G)^2(\text{Re}F)^2-4(\text{Re}F)^2(\text{Im}F)^2}\ \Big].
\end{equation}
The protocol success probability is therefore given by 
\begin{equation}
\label{eq:protocolsuccessprobabilityintermsofpairwiseoverlaps}
\begin{split}
P_{\text{1bit}}&=\frac{1}{2}\bigg[1+\frac{1}{4}\sqrt{(\lambda_0+\lambda_2)(\lambda_1+\lambda_3)+\sqrt{(\lambda_0+\lambda_2)^2(\lambda_1+\lambda_3)^2-16\lambda_0\lambda_1\lambda_2\lambda_3}}\\
&+\frac{1}{4}\sqrt{(\lambda_0+\lambda_2)(\lambda_1+\lambda_3)-\sqrt{(\lambda_0+\lambda_2)^2(\lambda_1+\lambda_3)^2-16\lambda_0\lambda_1\lambda_2\lambda_3}}\bigg]\\
&=\frac{1}{2}\bigg[1+\frac{1}{2}\sqrt{1-G^2+2\sqrt{(1+G)^2(\text{Im}F)^2+(1-G)^2(\text{Re}F)^2-4(\text{Re}F)^2(\text{Im}F)^2}}\\
&+\frac{1}{2}\sqrt{1-G^2-2\sqrt{(1+G)^2(\text{Im}F)^2+(1-G)^2(\text{Re}F)^2-4(\text{Re}F)^2(\text{Im}F)^2}}\bigg].
\end{split}
\end{equation}
We calculate $B_{\text{OT}}$ and $P_{\text{1bit}}$ for all the examples examined, both for pure and mixed states.
For the three-mode pure symmetric coherent states in \cref{eq:purethreemode} it holds that
\begin{equation} \label{eq:pairwiseoverlapsthreemode}
\begin{split}
F & =\bra{\alpha,\alpha,\alpha}\ket{\alpha,\alpha,-\alpha}=e^{-2\abs{\alpha}^2} \\
G & =\bra{\alpha,\alpha,\alpha}\ket{\alpha,-\alpha,-\alpha}=e^{-4\abs{\alpha}^2}=F^2.
\end{split}
\end{equation}
If four symmetric states are used for oblivious transfer, and Bob selects which bit he would like obtain, then the success probability is the minimum-error probability for distinguishing between the corresponding pairs of states. The protocol success probability in terms of $F$ and $G$ is
\begin{equation} \label{eq:protocolsuccessprobabilitypurethreemode}
\begin{split}
P_{\text{1bit}}^{\text{pure}} & =\frac{1}{2}\bigg[1\ +\ \frac{1}{2}\sqrt{1-G^2+2\abs{(1-G)F}}+\frac{1}{2}\sqrt{1-G^2-2\abs{(1-G)F}}\ \bigg] \\
& =\frac{1}{2}\bigg[1\ +\ \frac{1}{2}\sqrt{1-F^4+2F(1-F^2)}+\frac{1}{2}\sqrt{1-F^4-2F(1-F^2)}\ \bigg] \\
& =\frac{1}{2}\bigg[1\ +\ \frac{1}{2}\sqrt{(1-F)(1+F)^3}+\frac{1}{2}\sqrt{(1+F)(1-F)^3}\ \bigg]=\frac{1}{2}\Big[1+\sqrt{1-F^2}\ \Big].
\end{split}
\end{equation}
Bob's cheating probability is the minimum-error probability for distinguishing between all four states. In terms of $F$ and $G$, the cheating probability is
\begin{equation} \label{eq:cheatingprobabilitypurethreemode}
\begin{split}
B_{\text{OT}}^{\text{pure}} & =\frac{1}{16}\Big(\sqrt{1+G+2F}+\sqrt{1+G-2F}+2\sqrt{1-G}\ \Big)^2 \\
& =\frac{1}{16}\big(2+2\sqrt{1-F^2}\ \big)^2=\frac{1}{16}\big(4P_{\text{1bit}}^{\text{pure}}\big)^2=(P_{\text{1bit}}^{\text{pure}})^2.
\end{split}
\end{equation}
For the three-mode phase-randomized coherent states, Bob's cheating probability is 
\begin{equation} \label{eq:cheatingprobabilitymixedthreemode}
\begin{split}
B_{\text{OT}} & =\frac{e^{-3\abs{\alpha}^2}}{16}\sum\limits_{N=0}^{\infty}{\frac{(3\abs{\alpha}^2)^{N}}{N!}\Big(3\sqrt{1-G_N}+\sqrt{1+3G_N}\Big)^2} \\
& =\frac{e^{-3\abs{\alpha}^2}}{16}\sum\limits_{N=0}^{\infty}{\frac{(3\abs{\alpha}^2)^{N}}{N!}\Big(10-6G_N+6\sqrt{(1-G_N)(1+3G_N)}\ \Big)} \\
& =\frac{4}{16}+\frac{6}{4}\frac{e^{-3\abs{\alpha}^2}}{4}\sum\limits_{N=0}^{\infty}{\frac{(3\abs{\alpha}^2)^N}{N!}\Big(1-G_N+\sqrt{-3G_N^2+2G_N+1}\ \Big)} \\
& =\frac{1}{4}+\frac{3}{2}\Big(P_{\text{1bit}}-\frac{1}{2}\Big)=\frac{3}{2}P_{\text{1bit}}-\frac{1}{2}.
\end{split}
\end{equation}
For the four-mode pure symmetric coherent states in \cref{num3}, the pairwise overlaps are
\begin{equation} \label{eq:pairwiseoverlapsfourmode}
\begin{split}
\tilde{F} & =\bra{\alpha,\alpha,\alpha,-\alpha}\ket{\alpha,\alpha,-\alpha,\alpha}=e^{-4\abs{\alpha}^2} \\
\tilde{G} & =\bra{\alpha,\alpha,\alpha,-\alpha}\ket{\alpha,-\alpha,\alpha,\alpha}=e^{-4\abs{\alpha}^2}=\tilde{F}.
\end{split}
\end{equation}
The protocol success probability in terms of $F$ and $G$ is then
\begin{equation} \label{eq:protocolsuccessprobabilitypurefourmode}
\begin{split}
\tilde{P}_{\text{1bit}}^{\text{pure}} & =\frac{1}{2}\bigg[1\ +\ \frac{1}{2}\sqrt{1-\tilde{G}^2+2(1-\tilde{G})\tilde{F}}+\frac{1}{2}\sqrt{1-\tilde{G}^2-2(1-\tilde{G})\tilde{F}}\ \bigg] \\
& =\frac{1}{2}\bigg[1\ +\ \frac{1}{2}\sqrt{1-\tilde{F}^2+2\tilde{F}(1-\tilde{F})}+\frac{1}{2}\sqrt{1-\tilde{F}^2-2\tilde{F}(1-\tilde{F})}\ \bigg] \\
& =\frac{3-\tilde{F}}{4}+\frac{\sqrt{(1-\tilde{F})(1+3\tilde{F})}}{4}.
\end{split}
\end{equation}
The cheating probability for the pure four-mode coherent states is
\begin{equation} \label{eq:cheatingprobabilitypurefourmode}
\begin{split}
\tilde{B}_{\text{OT}}^{\text{pure}} & =\frac{1}{16}\big(\sqrt{1+3\tilde{F}}+3\sqrt{1-\tilde{F}}\ \big)^2=\frac{1}{16}\Big(10-6\tilde{F}+6\sqrt{(1-\tilde{F})(1+3\tilde{F})}\Big) \\
& =\frac{6}{16}\Big(3-\tilde{F}+\sqrt{(1-\tilde{F})(1+3\tilde{F})}\Big)-\frac{8}{16}=\frac{6}{16}\big(4\tilde{P}_{\text{1bit}}^{\text{pure}}\big)-\frac{8}{16}=\frac{3}{2}\tilde{P}_{\text{1bit}}^{\text{pure}}-\frac{1}{2}.
\end{split}
\end{equation}
For the four-mode phase-randomized coherent states, because $\tilde{F}_N=\tilde{G}_N=0$ for $N>0$, and $\tilde{F}_0=\tilde{G}_0=1$ for $N=0$, we have $\tilde{P}_{\text{1bit},N}=1$ and   $\tilde{P}_{\text{1bit},0}=\frac{1}{2}$. The protocol success probability is thus
\begin{equation}
\tilde{P}_{\text{1bit}}=\sum_{N=0}^\infty p_N\tilde{P}_{\text{1bit},N}=1-\frac{e^{-4\abs{\alpha}^2}}{2}.
\end{equation}
For the relation between $\tilde{B}_{\text{OT}}$ and $\tilde{P}_{\text{1bit}}$, we have
\begin{equation} \label{eq:cheatingprobabilitymixedfourmode}
\tilde{B}_{\text{OT}}=1-\frac{3e^{-4\abs{\alpha}^2}}{4}=1-\frac{3}{2}\big(1-\tilde{P}_{\text{1bit}}\big)=\frac{3}{2}\tilde{P}_{\text{1bit}}-\frac{1}{2}.
\end{equation}
We would like to see whether the phase-randomized states perform better or worse than their pure-state counterparts, without phase randomization. For the pure phase-encoded states in \cref{num4}, we have
\begin{equation} \label{eq:pairwiseoverlapsphaseencoded}
\begin{split}
\bar{F} & =\bra{\alpha,\alpha}\ket{\alpha,i\alpha}=e^{-\abs{\alpha}^2+i\abs{\alpha}^2}=e^{-\abs{\alpha}^2}(\cos{\abs{\alpha}^2}+i\sin{\abs{\alpha}^2}) \\
\bar{G} & =\bra{\alpha,\alpha}\ket{\alpha,-\alpha}=e^{-2\abs{\alpha}^2}.
\end{split}
\end{equation}
The protocol success probability in terms of $\bar{F}$ and $\bar{G}$ for the pure phase-encoded states is
\begin{equation} \label{eq:protocolsuccessprobabilitypurephaseencoded}
\begin{split}
\bar{P}_{\text{1bit}}^{\text{pure}} & =\frac{1}{2}\bigg[1\ +\ \frac{1}{2}\sqrt{1-\bar{G}^2+2\sqrt{(1+\bar{G})^2(\text{Im}\bar{F})^2+(1-\bar{G})^2(\text{Re}\bar{F})^2-4(\text{Re}\bar{F})^2(\text{Im}\bar{F})^2}} \\
& +\ \frac{1}{2}\sqrt{1-\bar{G}^2-2\sqrt{(1+\bar{G})^2(\text{Im}\bar{F})^2+(1-\bar{G})^2(\text{Re}\bar{F})^2-4(\text{Re}\bar{F})^2(\text{Im}\bar{F})^2}}\ \bigg] \\
& =\frac{1}{2}+\frac{1}{4}\sqrt{1-e^{-4\abs{\alpha}^2}+2\sqrt{e^{-6\abs{\alpha}^2}+e^{-2\abs{\alpha}^2}-e^{-4\abs{\alpha}^2}(2\cos{2\abs{\alpha}^2}+\sin^2{2\abs{\alpha}^2})}} \\
& +\frac{1}{4}\sqrt{1-e^{-4\abs{\alpha}^2}-2\sqrt{e^{-6\abs{\alpha}^2}+e^{-2\abs{\alpha}^2}-e^{-4\abs{\alpha}^2}(2\cos{2\abs{\alpha}^2}+\sin^2{2\abs{\alpha}^2})}} \\
& =\frac{1}{2}+\frac{1}{4}\sqrt{1-e^{-4\abs{\alpha}^2}+2e^{-2\abs{\alpha}^2}\sqrt{2\cosh{2\abs{\alpha}^2}-2\cos{2\abs{\alpha}^2}-\sin^2{2\abs{\alpha}^2}}} \\
& +\frac{1}{4}\sqrt{1-e^{-4\abs{\alpha}^2}-2e^{-2\abs{\alpha}^2}\sqrt{2\cosh{2\abs{\alpha}^2}-2\cos{2\abs{\alpha}^2}-\sin^2{2\abs{\alpha}^2}}} \\
& =\frac{1}{2}+\frac{e^{-\abs{\alpha}^2}}{2\sqrt{2}}\sqrt{\sinh{2\abs{\alpha}^2}+\sqrt{2\cosh{2\abs{\alpha}^2}-2\cos{2\abs{\alpha}^2}-\sin^2{2\abs{\alpha}^2}}} \\
& +\frac{e^{-\abs{\alpha}^2}}{2\sqrt{2}}\sqrt{\sinh{2\abs{\alpha}^2}-\sqrt{2\cosh{2\abs{\alpha}^2}-2\cos{2\abs{\alpha}^2}-\sin^2{2\abs{\alpha}^2}}} \\
& =\frac{1}{2}+\frac{e^{-\abs{\alpha}^2}}{2}\sqrt{\sinh{2\abs{\alpha}^2}+\sqrt{\sinh^2{2\abs{\alpha}^2}-2\cosh{2\abs{\alpha}^2}+2\cos{2\abs{\alpha}^2}+\sin^2{2\abs{\alpha}^2}}}.
\end{split}
\end{equation}
The cheating probability for the pure phase-encoded states in terms of $\bar{F}$ and $\bar{G}$ is
\begin{equation} \label{eq:cheatingprobabilitypurephaseencoded}
\begin{split}
\bar{B}_{\text{OT}}^{\text{pure}} & =\frac{1}{16}\big(\sqrt{\bar{\lambda}_1}+\sqrt{\bar{\lambda}_2}+\sqrt{\bar{\lambda}_3}+\sqrt{\bar{\lambda}_4}\ \big)^2 \\
& =\frac{1}{16}\Big(\sqrt{1+\bar{G}+2\text{Re}\bar{F}}+\sqrt{1+\bar{G}-2\text{Re}\bar{F}}+\sqrt{1-\bar{G}+2\text{Im}\bar{F}}+\sqrt{1-\bar{G}-2\text{Im}\bar{F}}\ \Big)^2 \\
& =\frac{1}{8}\Big(\sqrt{1+\bar{G}+\sqrt{(1+\bar{G})^2-4(\text{Re}\bar{F})^2}}+\sqrt{1-\bar{G}+\sqrt{(1-\bar{G})^2-4(\text{Im}\bar{F})^2}}\Big)^2 \\
& =\frac{1}{8}\bigg(\sqrt{1+e^{-2\abs{\alpha}^2}+\sqrt{(1+e^{-2\abs{\alpha}^2})^2-4e^{-2\abs{\alpha}^2}\cos^2{\abs{\alpha}^2}}} \\
& +\sqrt{1-e^{-2\abs{\alpha}^2}+\sqrt{(1-e^{-2\abs{\alpha}^2})^2-4e^{-2\abs{\alpha}^2}\sin^2{\abs{\alpha}^2}}}\ \bigg)^2 \\
& =\frac{e^{-\abs{\alpha}^2}}{4}\bigg(\sqrt{\cosh{\abs{\alpha}^2}+\sqrt{\cosh^2{\abs{\alpha}^2}-\cos^2{\abs{\alpha}^2}}}+\sqrt{\sinh{\abs{\alpha}^2}+\sqrt{\sinh^2{\abs{\alpha}^2}-\sin^2{\abs{\alpha}^2}}}\ \bigg)^2.
\end{split}
\end{equation}
The Gram matrix for the states $|\bar{\psi}_{bc,N}\rangle$ in eq. \eqref{eq:phaseencodedNpure} for $N>0$ is
\begin{equation} \label{eq:grammatrixphaseencodedNpositive} \bar{\mathcal{G}}_N=\begin{pmatrix}
1 & \Big(\frac{1+i}{2}\Big)^N & 0 & \Big(\frac{1-i}{2}\Big)^N \\
\Big(\frac{1-i}{2}\Big)^N & 1 & \Big(\frac{1+i}{2}\Big)^N & 0 \\
0 & \Big(\frac{1-i}{2}\Big)^N & 1 & \Big(\frac{1+i}{2}\Big)^N \\
\Big(\frac{1+i}{2}\Big)^N & 0 & \Big(\frac{1-i}{2}\Big)^N & 1
\end{pmatrix}.
\end{equation}
For $N=0$, the four states $|\bar\psi_{bc,0}\rangle$ are all equal to the vacuum state, and therefore indistinguishable. % is the same as in \cref{eq:grammatrixfourmodeNzero}, 
%\Erika{a matrix where every element is equal to $1$}.
Hence, for the phase-randomized phase-encoded states, because $\bar{G}_N=0$ and $\bar{F}_N=\Big(\frac{1+i}{2}\Big)^N$ for $N>0$, we have
\begin{equation} \label{eq:Nprotocolsuccessprobabilitymixedphaseencoded}
\begin{split}
\bar{P}_{\text{1bit},N} & =\frac{1}{2}\bigg[1\ +\ \frac{1}{2}\sqrt{1-\bar{G}_N^2+2\sqrt{(1+\bar{G}_N)^2(\text{Im}\bar{F}_N)^2+(1-\bar{G}_N)^2(\text{Re}\bar{F}_N)^2-4(\text{Re}\bar{F}_N)^2(\text{Im}\bar{F}_N)^2}} \\
& +\ \frac{1}{2}\sqrt{1-\bar{G}_N^2-2\sqrt{(1+\bar{G}_N)^2(\text{Im}\bar{F}_N)^2+(1-\bar{G}_N)^2(\text{Re}\bar{F}_N)^2-4(\text{Re}\bar{F}_N)^2(\text{Im}\bar{F}_N)^2}}\ \bigg] \\
& =\frac{1}{2}\bigg[1\ +\ \frac{1}{2}\sqrt{1+2\sqrt{(\text{Im}\bar{F}_N)^2+(\text{Re}\bar{F}_N)^2-4(\text{Re}\bar{F}_N)^2(\text{Im}\bar{F}_N)^2}} \\
& +\ \frac{1}{2}\sqrt{1-2\sqrt{(\text{Im}\bar{F}_N)^2+(\text{Re}\bar{F}_N)^2-4(\text{Re}\bar{F}_N)^2(\text{Im}\bar{F}_N)^2}}\ \bigg].
\end{split}
\end{equation}
We convert $\bar{F}_N$ from rectangular to polar form,
\begin{equation} \label{eq:Npairwiseoverlapsmixedphaseencoded}
\begin{split}
\bar{F}_N & =\Big(\frac{1+i}{2}\Big)^N=\Big(\frac{\sqrt{2}e^{i\pi/4}}{2}\Big)^N=2^{-N/2}e^{iN\pi/4}=2^{-N/2}\big[\cos{(N\pi/4)}+i\sin{(N\pi/4)}\big] \\
\bar{F}_N^* & =\Big(\frac{1-i}{2}\Big)^N=\Big(\frac{\sqrt{2}e^{-i\pi/4}}{2}\Big)^N=2^{-N/2}e^{-iN\pi/4}=2^{-N/2}\big[\cos{(N\pi/4)}-i\sin{(N\pi/4)}\big].
\end{split}
\end{equation}
We find that
\begin{equation} \label{eq:expressionsNpairwiseoverlapsmixedphaseencoded}
\begin{split}
(\text{Re}\bar{F}_N)^2+(\text{Im}\bar{F}_N)^2 & =2^{-N}\cos^2{(N\pi/4)}+2^{-N}\sin^2{(N\pi/4)}=2^{-N} \\
4(\text{Re}\bar{F}_N)^2(\text{Im}\bar{F}_N)^2 & =4^{-N}4\cos^2{(N\pi/4)}\sin^2{(N\pi/4)}=4^{-N}\sin^2{(N\pi/2)}.
\end{split}
\end{equation}
For $N=0$ we have $\bar{F}_0=\bar{G}_0=1$ and $\bar{P}_{\text{1bit},0}=\frac{1}{2}$, so the total protocol success probability becomes
\begin{equation} \label{eq:protocolsuccessprobabilitymixedphaseencoded}
\begin{split}
\bar{P}_{\text{1bit}} & =\sum_{N=0}^{\infty}\bar{p}_N\bar{P}_{\text{1bit},N}=\bar{p}_0\bar{P}_{\text{1bit},0}+\sum_{N=1}^{\infty}\bar{p}_N\bar{P}_{\text{1bit},N}=\frac{e^{-2\abs{\alpha}^2}}{2}+\frac{e^{-2\abs{\alpha}^2}}{2}\sum_{N=1}^{\infty}{\frac{(2\abs{\alpha}^2)^N}{N!}} \\
& +\ \frac{e^{-2\abs{\alpha}^2}}{4}\sum_{N=1}^{\infty}{\frac{(2\abs{\alpha}^2)^N}{N!}\Bigg(\sqrt{1+2\sqrt{2^{-N}-4^{-N}\sin^2{(N\pi/2)}}}}\ +\ \sqrt{1-2\sqrt{2^{-N}-4^{-N}\sin^2{(N\pi/2)}}}\ \Bigg) \\
& =\frac{1}{2}\ +\ \frac{e^{-2\abs{\alpha}^2}}{4}\sum_{N=1}^{\infty}{\frac{(2\abs{\alpha}^2)^N}{N!}\Bigg(\sqrt{1+2\sqrt{2^{-N}-4^{-N}\sin^2{(N\pi/2)}}}}\ +\ \sqrt{1-2\sqrt{2^{-N}-4^{-N}\sin^2{(N\pi/2)}}}\ \Bigg).
\end{split}
\end{equation}
To obtain a simpler form for the protocol success probability, we note that
\begin{equation} \label{eq:sumtwosquarerootstrick}
\big(\sqrt{A+B}+\sqrt{A-B}\big)^2=2A+2\sqrt{A^2-B^2}\Rightarrow\sqrt{A+B}+\sqrt{A-B}=\pm\sqrt{2(A+\sqrt{A^2-B^2})}
\end{equation}
where we accept the positive solution. For $A=1$ and $B= 2\sqrt{2^{-N}-4^{-N}\sin^2{(N\pi/2)}}$,
\begin{equation} \label{eq:usetrickforphaseencodedstates}
A^2-B^2=1-4\big(2^{-N}-4^{-N}\sin^2{(N\pi/2)}\big),
\end{equation}
so the protocol success probability is
\begin{equation} \label{eq:finalprotocolsuccessprobabilitymixedphaseencoded}
\bar{P}_{\text{1bit}}=\frac{1}{2}+\frac{e^{-2\abs{\alpha}^2}}{2\sqrt{2}}\sum_{N=1}^{\infty}{\frac{(2\abs{\alpha}^2)^N}{N!}\sqrt{1+\sqrt{4^{-N+1}\sin^2{(N\pi/2)}-2^{-N+2}+1}}}.
\end{equation}
For the respective cheating probability when we run the protocol with the phase-randomized phase-encoded states, the eigenvalues of the Gram matrix in \cref{eq:grammatrixphaseencodedNpositive} are $\bar{\lambda}_{0,2}^{(N)}=1\pm\Big[\Big(\frac{1+i}{2}\Big)^N+\Big(\frac{1-i}{2}\Big)^N\Big]$ and $\bar{\lambda}_{1,3}^{(N)}=1\pm i\Big[\Big(\frac{1+i}{2}\Big)^N-\Big(\frac{1-i}{2}\Big)^N\Big]$ for $N>0$ and $\bar{\lambda}_0^{(0)}=4$, $\bar{\lambda}_{1,2,3}^{(0)}=0$ for $N=0$. The optimal cheating probability when using these states is 
\begin{equation} \label{eq:cheatingprobabilitymixedphaseencoded}
\begin{split}
\bar{B}_{\text{OT}} & =e^{-2\abs{\alpha}^2}\sum_{N=0}^{\infty}{\frac{\abs{\alpha}^{2N}\bar{M}_N}{16}\bigg(\sqrt{\bar{\lambda}_0^{(N)}}+\sqrt{\bar{\lambda}_1^{(N)}}+\sqrt{\bar{\lambda}_2^{(N)}}+\sqrt{\bar{\lambda}_3^{(N)}}\bigg)^2} \\
& =\frac{e^{-2\abs{\alpha}^2}}{4}+\frac{e^{-2\abs{\alpha}^2}}{16}\sum_{N=1}^{\infty}{\frac{(2\abs{\alpha}^2)^N}{N!}\Big(\sqrt{\bar{\lambda}_0^{(N)}}+\sqrt{\bar{\lambda}_1^{(N)}}+\sqrt{\bar{\lambda}_2^{(N)}}+\sqrt{\bar{\lambda}_3^{(N)}}\Big)^2}.
\end{split}
\end{equation}
The eigenvalues of the Gram matrix for $N>0$ become
\begin{equation} \label{eq:Ngrammatrixeigenvaluesmixedphaseencoded}
\begin{split}
\bar{\lambda}_{0,2}^{(N)} & =1\pm\big(\bar{F}_N+\bar{F}_N^*\big)=1\pm2^{-N/2+1}\cos{(N\pi/4)} \\
\bar{\lambda}_{1,3}^{(N)} & =1\pm i\big(\bar{F}_N-\bar{F}_N^*\big)=1\mp2^{-N/2+1}\sin{(N\pi/4)},
\end{split}
\end{equation}
so the cheating probability can also be written as
\begin{equation} \label{eq:secondexpressioncheatingprobabilitymixedphaseencoded}
\begin{split}
\bar{B}_{\text{OT}} & =\frac{e^{-2\abs{\alpha}^2}}{4}+\frac{e^{-2\abs{\alpha}^2}}{16}\sum_{N=1}^{\infty}{\frac{(2\abs{\alpha}^2)^N}{N!}\bigg(\sqrt{1+2^{-N/2+1}\cos{(N\pi/4)}}} \\
& \ +\ \sqrt{1-2^{-N/2+1}\cos{(N\pi/4)}}+\sqrt{1+2^{-N/2+1}\sin{(N\pi/4)}}+\sqrt{1-2^{-N/2+1}\sin{(N\pi/4)}}\ \bigg)^2.
\end{split}
\end{equation}
To simplify this, we note that
\begin{equation} \label{eq:secondusetrickphaseencoded}
\begin{split}
\sqrt{1+2^{-N/2+1}\cos{(N\pi/4)}}\ +\ \sqrt{1-2^{-N/2+1}\cos{(N\pi/4)}} & =\sqrt{2\Big(1+\sqrt{1-2^{-N+2}\cos^2{(N\pi/4)}}\Big)} \\
\sqrt{1+2^{-N/2+1}\sin{(N\pi/4)}}\ +\ \sqrt{1-2^{-N/2+1}\sin{(N\pi/4)}} & =\sqrt{2\Big(1+\sqrt{1-2^{-N+2}\sin^2{(N\pi/4)}}\Big)}.
\end{split}
\end{equation}
So if we call $\mathcal{S}_N$ the quantity in the parenthesis inside the sum, we obtain
\begin{equation} \label{eq:snquantityfromtrickuse}
\mathcal{S}_N^2=2\bigg(\sqrt{1+\sqrt{1-2^{-N+2}\cos^2{(N\pi/4)}}}+\sqrt{1+\sqrt{1-2^{-N+2}\sin^2{(N\pi/4)}}}\bigg)^2.
\end{equation}
Therefore
\begin{equation} \label{eq:finalcheatingprobabilitymixedphaseencoded}
\begin{split} 
\bar{B}_\text{OT}=\frac{e^{-2\abs{\alpha}^2}}{4}+\frac{e^{-2\abs{\alpha}^2}}{8}\sum\limits_{N=1}^{\infty}\frac{(2\abs{\alpha}^2)^N}{N!}&\bigg[\sqrt{1+\sqrt{1-2^{-N+2}\cos^2{(N\pi/4)}}} \\
& +\sqrt{1+\sqrt{1-2^{-N+2}\sin^2{(N\pi/4)}}}\bigg]^2.
\end{split}
\end{equation}
The cheating probabilities for the corresponding pure and mixed sets of states are plotted in \cref{fig2,fig4,fig5} in the main paper. It is evident that mixing lowers the cheating probability for each value of $\abs{\alpha}$. It is apparent from the plots that mixing alters the $\abs{\alpha}$-dependence, resulting in a smoother shape.
\end{widetext}

\end{document}